\documentclass[useAMS]{mn2e}
\usepackage{amsmath}
\usepackage{url}
\usepackage{amsfonts}
\usepackage{amsbsy}
\usepackage{subfigure}
\usepackage{verbatim}
\usepackage{amssymb}
\usepackage{amsbsy}
\usepackage{graphicx}
\usepackage{array}
\usepackage{color}

\title[The stress-pressure relationship in the MRI]{The
  stress-pressure relationship in simulations of MRI-induced turbulence}
\author[Ross, Latter, \& Guilet]{Johnathan Ross$^{1}$\thanks{E-mail:
   jpjr2@cam.ac.uk},
   Henrik N. Latter$^{1}$,
   Jerome Guilet$^{1,2}$ \\
$^{1}$ DAMTP, University of Cambridge, CMS, Wilberforce Road,
Cambridge CB3 0WA, UK\\
$^{2}$ Max-Planck-Institut fur Astrophysik, Karl-Schwarzschild-Str. 1, D-85748 Garching, Germany  }
\date{}

\renewcommand{\thetable}{\Roman{table}}

\begin{document}

\maketitle

\begin{abstract}

We determine how MRI-turbulent stresses depend on gas pressure
via a suite of
unstratified shearing box simulations. Earlier numerical work reported
only a very weak dependence at best, results
that call into question the canonical $\alpha$-disk model and the
thermal stability results that follow from it. 
Our simulations, in contrast, exhibit
a stronger relationship, and show that
previous work was box-size limited:
turbulent `eddies' were artificially restricted by the numerical
domain
rather than
by the scale height.
Zero-net-flux runs without physical diffusion coefficients yield a
stress proportional to $P^{0.5}$, where $P$
is pressure. The stresses are also proportional to the grid length and
hence remain numerically unconverged.
The same runs with physical
diffusivities, however, give a result closer to an
$\alpha$-disk: the stress is $\propto P^{0.9}$. 
Net-flux simulations without explicit diffusion
exhibit stresses $\propto P^{0.5}$, but 
stronger imposed fields weaken this correlation.
In summary, compressibility is important for the saturation of the
MRI, but the exact stress-pressure relationship is difficult to
ascertain in local simulations 
because of numerical convergence issues and the influence of any imposed flux. 
As a consequence, the interpretation of thermal stability behaviour in
local simulations is a problematic enterprise.
\end{abstract}


\section{Introduction}

The accretion of gas through a disk, and ultimately on to a star or
black hole, powers the intense luminosity of a great many astrophysical
objects. The classical theory of disk accretion 
assumes that (a) correlated turbulent motions in the disk
apply a torque that drives the observed
transport, and that (b) the resulting radial-azimuthal component of the
stress $\Pi_{r\phi}$ is proportional to the gas pressure $P$, i.e.\
$\Pi_{r\phi}=\alpha P$ (Lynden-Bell \& Pringle 1973, Shakura \&
Sunyaev 1973). This model permits the closure of the
system of governing equations, allowing researchers to construct
disk solutions with which to interpret observations. 

At present the consensus is that the magnetorotational instability
(MRI) generates
disk turbulence, at
least in disks that are sufficiently ionised (Balbus \& Hawley 1991,
1998). Numerical simulations of the MRI in unstratified local domains certainly yield
appropriate values for $\alpha$ in cases where the computational
domain is penetrated by a strong magnetic field (Hawley et al.~1995,
Simon et al.~2009). 
In contrast to the measurement of $\alpha$, however, there have
been relatively few attempts to test whether
$\Pi_{r\phi}$ is in fact proportional to $P$. 
 Four studies exist: Hawley et al.~(1995, hereafter HGB95), Sano et
al.~(2004, SITS04), Simon
et al.~(2009, SHB09) and
most recently Minoshima et al.~(2015, MHS15) which 
appeared when this paper was in draft form. All four were
undertaken in unstratified shearing boxes, 
and show that $\Pi_{r\phi}$ 
depends on $P$ to a very weak power or not at all (see also Blackman
et al.~2008). 
Taken on face value, these results imply that the MRI saturates
with little or no recourse to compressibility, and moreover cast
doubt on the validity of the $\alpha$-model, and the many structure
and stability results that issue from it.

In this paper, we re-examine the relationship between the MRI-induced
turbulent stress and the gas pressure with
numerical simulations in local unstratified boxes.
We employ the codes Ramses (Teyssier 2002, Fromang et
al.~2006). 
 In our main runs the gas is
permitted to heat up, via turbulent dissipation, and we compare the
correlation between $\Pi_{r\phi}$ and $P$ during this phase.
Special care has been taken to minimise the influence of the box size on our
results, and so we have set $H<L$ in most runs, where $H$ is the `scale
height' (the characteristic distance travelled by a sound wave over
one orbit) and $L$ is the vertical and radial box lengths. This is key. In the
opposite regime, $H>L$, the turbulent eddies
 are always limited by the box (a numerical effect) and not
 by compressibility (a physical effect). As a consequence, the
 relationship between MRI saturation and compressibility is lost.
Note that the previous simulations of HGB95, 
SITS04, SHB09, and MHS15 almost always use $H>L$, and hence
suffer from this shortcoming.

Runs with $L \geq 2H$, yield a 
$\Pi_{r\phi}$ that depends on $P$ in a stronger way than in earlier
work.
When there is no net magnetic flux and no explicit diffusion
coefficients, we obtain $\Pi_{r\phi}\propto P^q$ 
with $q\approx 0.4-0.6$ in both heating and cooling runs. In fact, the
simulations indicate that $\Pi_{r\phi}\sim H\Delta$, where $\Delta$
is the grid length. This result shows that the 
stress is unconverged with respect to the numerical parameters --- a pathology of this particular
set-up (see also Fromang \& Papaloizou 2007).     
A suite of isothermal simulations
of differing box sizes and differing resolutions confirms this
basic idea.
The incorporation of physical diffusivities, however, elicits strikingly
different behaviour: the stress and pressure are then
almost proportional, with
$q\approx 0.9$. 
On the other hand, both net-toroidal and net-vertical flux simulations yield
$q\approx0.2-0.5$. 
In these cases $q$ depends on the strength of
the background magnetic field. The stronger the imposed flux, the
weaker the correlation between $\Pi_{r\phi}$ and $P$. 

Despite these various complications, we have demonstrated
that compressibility is important in the saturation of the
MRI. The exact form of the stress-pressure relationship, however, is
difficult to extract from local simulations because of the
issue of numerical convergence and the strength of any imposed
flux. This makes local simulations of thermal stability particularly difficult
to interpret.

The structure of the paper is as follows. In the next two sections we
outline a very basic theoretical framework with which to
interpret our results, and then give details of 
the numerical model and methods with which we attack the problem. Our
results appear in Sections 4 and 5, which treat zero-net-flux and
net-flux configurations separately.
 We bring everything together in Section 6 and discuss 
implications for the saturation of the MRI
and the possibility of thermal instability.

\section{Theoretical expectations}

In this section we restate the heuristic, and essentially
hydrodynamical, arguments that justify the $\alpha$-prescription. We
then show how an insufficiently large box impinges on the stress's
dependence on $P$. Finally, we speculate on how the presence of 
magnetic fields may change this picture.

\subsection{Hydrodynamical arguments}

Our initial assumption is that shear turbulence will act so as
to eradicate the destabilising conditions from which it sprung. In
other words, it will transport as much angular momentum outward
as is possible. We next assume that the only restriction
on the efficiency of this transport comes from the finite thickness of
the disk and from compressibility: turbulent eddies cannot be larger than
$H$, and turbulent speeds cannot exceed the sound speed $c_s$. If the
motions were faster, enhanced dissipation from shocks would
slow them till they were subsonic. In order to maximise
transport, however, the turbulence will induce
flows as close to $c_s$ as it can. Note that we are assuming there
is not a more stringent restriction, arising from a separate
incompressible mechanism, that limits the turbulence to shorter scales.
Consequently, we may write
\begin{align*}
\Pi_{r\phi} \sim \rho\, v_{r} v_{\phi} \sim \rho\,
(l_\text{turb}\Omega)^2,
\end{align*}
where $v_{r}$ and $v_{\phi}$ are the characteristic radial and
azimuthal speeds of
the largest eddies, and $l_\text{turb}$ is their characteristic size.
Letting either $v \lesssim c_s$ or $l_\text{turb} \lesssim
H$ yields the alpha prescription,
and we have $\Pi_{r\phi} =\alpha P$, for some constant $\alpha<1$. 
In terms of the scale height, $\Pi_{r\phi}\sim H^2$. 

An alternative argument uses dimensional analysis, a version of which
we now briefly give. Imagine a `perfect' shearing box simulation, perfect in the
sense that its outcome is independent of the numerical domain size or the
small scales. There are three relevant physical quantities, $\Pi_{r\phi}$, $c_s$,
$\rho$ (whose dependence on time is implicit), and two physical dimensions, mass density, and speed. 
There is hence only
one way to relate these quantities, $\Pi_{r\phi} \propto \rho c_s^2$,
and we arrive at the alpha model once again\footnote{Note that if
viscous diffusion was explicitly included in the simulation (and
deemed important) then the constant of proportionality would become a
function of Reynolds number and the dependence on $c_s$ may be more complicated.}.

\subsection{Box-size limitations}

Consider numerical simulations performed in unstratified boxes
with a vertical and radial size of $L$. If $L \gg H$, then the
computational domain should play a negligible role,
and the above
argument will hold: $\Pi_{r\phi} \sim H^2$. 
 If, on the other hand, $L \ll H$
then the turbulent eddies
will be limited not by $H$ but by $L$. With $l_\text{eddy} \lesssim
L$, one obtains $\Pi_{r\phi} \sim L^2$, and the stress becomes a
constant, independent of $H$
(and hence $P$). This could be considered the `incompressible limit'. 
When $L\sim H$
these two limits should smoothly join up, and in Fig.~1 we provide 
a sketch showing how this might look.

\begin{figure}
\begin{center}
\scalebox{0.475}{\includegraphics{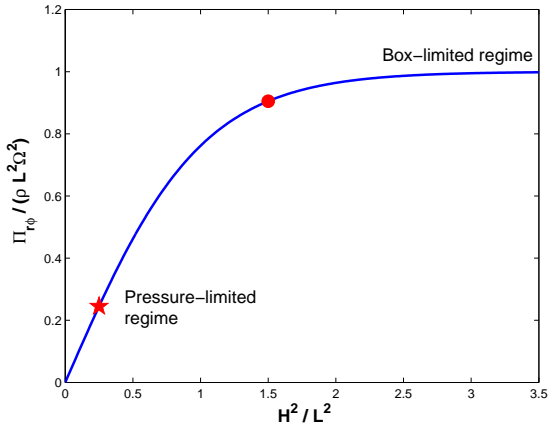}}
 \caption{A schematic graph of how we expect the stress
   $\Pi_{r\phi}$ to depend
   on scale height $H$ in an unstratified MRI simulation with a box
   size of $L$. In
   monotonically heating
   runs the stress will evolve by following the blue curve from left
   to right. Very roughly, the box-limited regime is to the right of the red dot, whereas
 the pressure-limited regime is to the left of the red star. Ideally,
 simulations should begin to the left of the red star.}
\label{disp}
\end{center}
\end{figure}

The simulations of HGB95, SITS04 and SHB09 allow
the box to heat up due to turbulent dissipation. Hence the scale height $H$
increases monotonically over time, reaching values, in some cases, many
orders of magnitude greater than its starting value $H_0$. In Fig.~1
we may treat $H^2$ as a proxy for time, with the evolution of the stress
following the blue curve to the right, rising with $H$ and then
plateauing once the eddies hit the box size.
Importantly, $H_0>L$ in all the SHB09 runs and
in all but a handful of the SITS04 and MHS15 runs. 
Thus their simulations' initial states 
fall mainly to the right of the red dot in Fig.~1. 
This means that the stress's
evolution is strongly constrained by the box, capable of
 increasing marginally, if at all.
 As a consequence, these studies are unable to truly test how strongly
 $\Pi_{r\phi}$ depends on $P$. 

To remove the
artificial effect of the box size we must begin to the
left of the red dot, ideally to the left of the red star.
The main aim of this paper is to present runs with initial states as
deep into this regime as possible.

\subsection{Magnetohydrodynamical complications}

MRI turbulence
involves magnetic fields, obviously, which we expect to spoil this 
attractively simple picture. For starters, simulations without a net
flux highlight the smallest scales over the largest. In the absence of
physical diffusivities, the characteristic eddy scale 
$l_\text{turb}$ prefers to sit near the fixed grid
scale $\Delta$, and as a result simulations return $\Pi_{r\phi}\propto \Delta$ (Fromang \& Papaloizou
2007, Simon et al.~2009). 
This dependence of the stress on $\Delta$ may partly wash out the dependence on $H$.

When physical
diffusivities are incorporated the situation is not much improved: the
ensuing turbulent dynamo depends on the magnetic Prandtl number, and
thus on the (fixed) small scales (Fromang et al.~2007, Riols et al.~2013,
2015). Having said that, large-scales are not completely irrelevant. 
In boxes
of different aspect ratios the dynamo exhibits
long term oscillations with coherent magnetic field reversals on the
box size (Lesur \& Ogilvie 2008a, b). Evidently, the dynamics are
complicated and may involve interactions between multiple scales. 
Certainly, the stress-pressure relationship could differ from the
picture described in Section 2.1.

On the other hand, simulations involving a net flux
witness the excitation of large-scale coherent motions,
such as channel flows. Our simple picture of transport via
turbulent eddies may be complicated by these features, which
periodically emerge
from the turbulent melee and dominate the angular momentum transport
(HGB95, Inutsuka \& Sano 2001, Bodo et al.~2008, SHB09).
Compressibility fails to limit the development of channel
flows, instead concentrating them into thin jets and
current sheets (Latter et al.~2009, Lesaffre et al.~2009). If these
flows dominate $\Pi_{r\phi}$ it is likely that the $H$ dependence
explored above is muddled or completely lost. 

More generally, turbulent transport in MHD is dominated by
 an anisotropic
tangle of flux
linkages which take time to develop and also to destroy. How
these magnetic structures respond as the pressure rises and falls is
unclear, but
 appreciable time lags may build up
between the variations in pressure and in magnetic transport, involving no doubt
the efficiency of magnetic reconnection. As a consequence, the relationship between
stress and pressure may not be a simple power law, but may include
memory effects, for example (cf.\ Ogilvie 2003, Pessah et al.~2006).

Though not an MHD effect per se, but one that relates
to time lags, is the issue of
causation: is $P$ driving $\Pi_{r\phi}$ or vice versa?
Section 2.1 argues that as pressure increases, the stress
follows. But a variation in stress can also force a change in the
temperature (and hence pressure) due to 
its associated variation in dissipation (see arguments in Hirose et
al.~2009). Though this interdependence
 is undoubtedly a complication, the timescales of the two
 processes differ and can be partly separated;
 the stress's action on the pressure
 occurs on shorter timescales than the pressure's action on
 the stress (Latter \& Papaloizou 2012). 
 
We have flagged quite a number of issues in this subsection, mainly
for reference.
In practice, not all directly impinge on our results. 
In what follows we explore primarily the
significance of the small scales and the strength of any net flux. 
But before we show our results we present the details of our physical
model and numerical set-up.

\section{Governing equations and numerical set up}

In this paper we solve the equations of compressible MHD in the
unstratified shearing box approximation (Goldreich and Lynden-Bell
1965). It uses a local Cartesian frame of reference which is
corotating with
a Keplerian disk at some arbitrary radius, $R_{0}$, with angular frequency $\boldsymbol{\Omega}=\Omega\boldsymbol{\hat{e}}_z$. As is conventional, $\boldsymbol{\hat{e}}_{x}$, $\boldsymbol{\hat{e}}_{y}$, $\boldsymbol{\hat{e}}_{z}$ are taken to be the unit vectors in the radial, azimuthal and vertical directions respectively. In this frame of reference, the equations of motion can be written as

\begin{align}
&\frac{\partial \rho}{ \partial t} + \nabla \cdot ( \rho \boldsymbol{v}
) = 0, \\
&\rho \frac{\partial \boldsymbol{v}}{\partial t} +\rho (\boldsymbol{v}
\cdot \nabla)\boldsymbol{v} 
 = - 2 \rho \bold{\Omega} \times \bold{v}  +
3\rho \Omega^{2}\boldsymbol{\hat{e}}_{x}-\nabla P \notag\\
& \hskip4cm  +( \nabla \times \boldsymbol{B} ) \times \boldsymbol{B} + \nabla\cdot\mathbf{T}, \\
&\frac{\partial \boldsymbol{B}}{\partial t} = \nabla \times
(\boldsymbol{v} \times \boldsymbol{B}) + \eta\nabla^2\boldsymbol{B},
\end{align}
where $\rho$ is the mass density, $\boldsymbol{v}$ is the velocity, $P$ is the gas pressure
$\boldsymbol{B}$ is the magnetic field, and $\eta$ is the magnetic
diffusivity. 
The (molecular) viscous stress is given by 
$\mathbf{T}= \nu(\nabla\boldsymbol{v} + \nabla\boldsymbol{v}^T)$, with
$\nu$ the shear viscosity. In most runs, $\nu=\eta=0$. 

We adopt either an isothermal or an ideal gas equation of state. In the
former 
\begin{equation}
P=\rho c_{0}^{2},
\end{equation} 
where $c_{0}$ is the fixed isothermal sound speed. Otherwise, we must
solve for the internal energy $\varepsilon$,
\begin{equation}
\frac{\partial \varepsilon}{\partial t} + \boldsymbol{v}\cdot\nabla
\varepsilon= -P\nabla\cdot\boldsymbol{v} +Q -\Lambda,
\end{equation}
 where $Q=\rho\nu |\nabla\times\boldsymbol{v}|^2 + \eta|\nabla\times\boldsymbol{B}|^2$ is the sum of the viscous heating and the resistive heating and
$\Lambda$ is a cooling function. For an ideal gas
\begin{equation}
\varepsilon= P/(\gamma-1),
\end{equation}
where $\gamma$ is the adiabatic index. We typically
take $\gamma= 7/5$. 
In non-isothermal runs, 
the sound speed is then given by $c_{s}=(\gamma P/\rho)^{1/2}$ and the
pressure scale height by $H=(2/\gamma)^{1/2}c_{s}/\Omega$. 
Finally, when non-zero, the cooling law is usually 
\begin{equation}
\Lambda = \theta\,P^{m},
\end{equation}
where $\theta$ and $m$ are constants. Some runs, however, set $\Lambda$
to be some fixed fraction of the dissipated energy.

\subsection{Numerical methods}

The set of equations just described are solved using Ramses, a finite
volume code based on the MUSCL-Hancock algorithm 
(Teyssier 2002; Fromang et al. 2006). Our version of the code solves
the shearing box equations on a uniform grid, 
and has been tested with an isothermal equation of state in
Fromang \& Stone (2009), Latter et al. (2010) and Fromang et
al. (2013).

Instead of the total $y$-momentum equation, we evolve the equivalent
conservation law for the angular momentum fluctuation $\rho
v_y^{\prime} =\rho( v_y - v_K)$,
 with $v_K$ the Keplerian velocity. The azimuthal advection arising
 from $v_K$ is solved using an upwind solver.
 Shearing box source terms in the momentum equation (due to tidal
 gravity and Coriolis forces) 
are implemented following the Crank-Nicholson algorithm described in
Stone \& Gardiner (2010). 

 The algorithm solves for the fluctuation energy $E^{\prime} \equiv P/(\gamma -1) + \rho
v^{\prime 2}/2 + B^2/2 $. In the absence of explicit dissipation, its conservation law is written as
\begin{equation} \label{3.1.1}
\frac{\partial E^\prime}{\partial t} +  \nabla\cdot \left(E^\prime{
    \boldsymbol{v}^\prime} + {\boldsymbol{v}^\prime} \cdot \mathbf{P} \right) = -v_K\frac{\partial E^\prime}{\partial y} + \left(B_xB_y - \rho v_x v_y^{\prime} \right)\frac{\partial v_K}{\partial x},
\end{equation}
where $\mathbf{P}$ is the total pressure tensor
\begin{equation}
\mathbf{P} = (P + B^2/2){\bf I} - \boldsymbol{B}\boldsymbol{B}.
\end{equation}
The left hand side Eq.~\eqref{3.1.1} is the usual energy conservation law,
which we solve using the MUSCL-Hancock algorithm. 
The treatment of the two terms on the right hand
have been modified: the azimuthal 
advection of energy is solved with an upwind solver, and the second
term involving the Maxwell and 
Reynolds stresses is added as a source term. 
Several numerical tests of this implementation are presented in
Appendix~A. 
The simulations presented in this paper used the HLLD Riemann solver
(Miyoshi \& Kusano 2005),
 and the multidimensional slope limiter described in Suresh (2000).

\subsection{Parameters and initial conditions}

We adopt the same units as HGB95, so that 
$\Omega =10^{-3}$,
the initial
density is $\rho_0=1$, and the initial sound speed $c_{s0}=10^{-3}$ in
diabatic runs, or $c_0=10^{-3}$ in isothermal runs.
Thus the initial scale height is $H_{0}=1$, though often we retain the
notation explicitly for clarity. Note that, in contrast,
$H$ is a function of $P$ and thus changes
in thermally evolving simulations, increasing as the box heats up, and
decreasing as it cools down. Finally, we denote by
$T_\text{orb}$ the period of one orbit.

Three initial configurations of magnetic field are considered: (a) zero
net-flux, for which $\boldsymbol{B}=B_{0} \sin (2 \pi 
x)\boldsymbol{\hat{e}}_{z} $,
(b) net-toroidal flux,
 $\boldsymbol{B}=B_{0}\boldsymbol{\hat{e}}_{y}$,
and (c)
 net-vertical flux, $\boldsymbol{B}=B_{0}\boldsymbol{\hat{e}}_{z}$.
 We define a plasma beta in code units through
$\beta=2/B_0^2$, which we set to $10^{3}$ unless otherwise stated. 
To induce the MRI we introduce random velocity perturbations in all
principle directions with amplitudes $ < 0.1c_{s0}$.
  
Typically the radial and vertical sizes of the computational domain
($L_x$ and $L_z$) are
the same and denoted by $L$, some multiple of $H_0$.
The azimuthal size is $L_{y}=5H_{0}$.
Unless otherwise stated, the resolution for the thermally evolving
simulations is $\Delta= H_{0}/N=1/64$, and is the same in all directions.

Physical diffusion is neglected in all but a handful of zero-net-flux
simulations 
for which
 $\nu=8\times 10^{-7}$ and $\eta=2\times 10^{-7}$. These values correspond to a magnetic Prandtl
 number of $P_m= \nu/\eta=4$, and Reynolds and magnetic Reynolds
 numbers of
 $Re \equiv H_0 c_{s0}/\nu= 1250$ and $R_m\equiv H_0 c_{s0}/\eta=5000$. These guarantee sustained turbulence and
 converged results (Fromang 2010).

\subsection{Diagnostics}

The transport of angular momentum is dictated by the turbulent stress which is given by the sum of the Reynolds and Maxwell stresses
\begin{equation}
\Pi_{xy}=-B_{x}B_{y}+\rho v_{x} v'_{y}.
\end{equation}
It is the behaviour of this quantity as $P$ varies that we are most
interested in. 
During diabatic simulations we calculate
$\langle \Pi_{xy}\rangle/P_0$, where the angled brackets indicate an instantaneous
box average and $P_0$ is the initial pressure. In isothermal
simulations we are interested in the time and box averaged rate of angular momentum transport, which corresponds to the usual definition of the alpha parameter: 
\begin{equation}
\alpha=\langle\langle\Pi_{xy}\rangle\rangle/P_{0}.
\end{equation}
The double angle
brackets represent averages over both volume and time, the latter
taken once we
judge the system to have entered its saturated state. 

In order to quantify the relationship between $\Pi_{xy}$ and $P$ we
assume that 
\begin{equation}
\Pi_{xy} \propto P^{q},
\end{equation}
for some number $q$ which we must determine. In heating runs, the pressure
increases monotonically, while the stress increases for some period of
time and then plateaus once the box size intervenes (in accord with
Fig.~1). 
By plotting the log of $\Pi_{xy}$
versus the log of $P$ during the growth phase we may obtain $q$. Unfortunately, the
calculation of $q$ is not unambiguous. The stress is often bursty and
the time of the growth phase relatively short. We hence can only
give a rough estimate for $q$.  

Finally, a useful diagnostic used by Lesur and Longaretti (2007) is the
vertical correlation length.
We define this correlation length by
\begin{align}
\hspace{-1cm}
& \zeta_{z}(v_{z})= \nonumber \\ 
&\bigg\langle\bigg\langle \frac{\int \int  v_{z}(x,y=L_{y}/2,z)v_{z}(x,y=L_{y}/2,z')dz'dz}{\int v^{2}_{z}(x,y=L_{y}/2,z)dz} \bigg\rangle\bigg\rangle_{T_\text{orb}}
\label{eq::cor}
\end{align} 
where the inner angled brackets represent an average over the $x$ direction and the outer angles brackets signify an average over 1 orbit.

\begin{figure*}
\includegraphics[width=8cm]{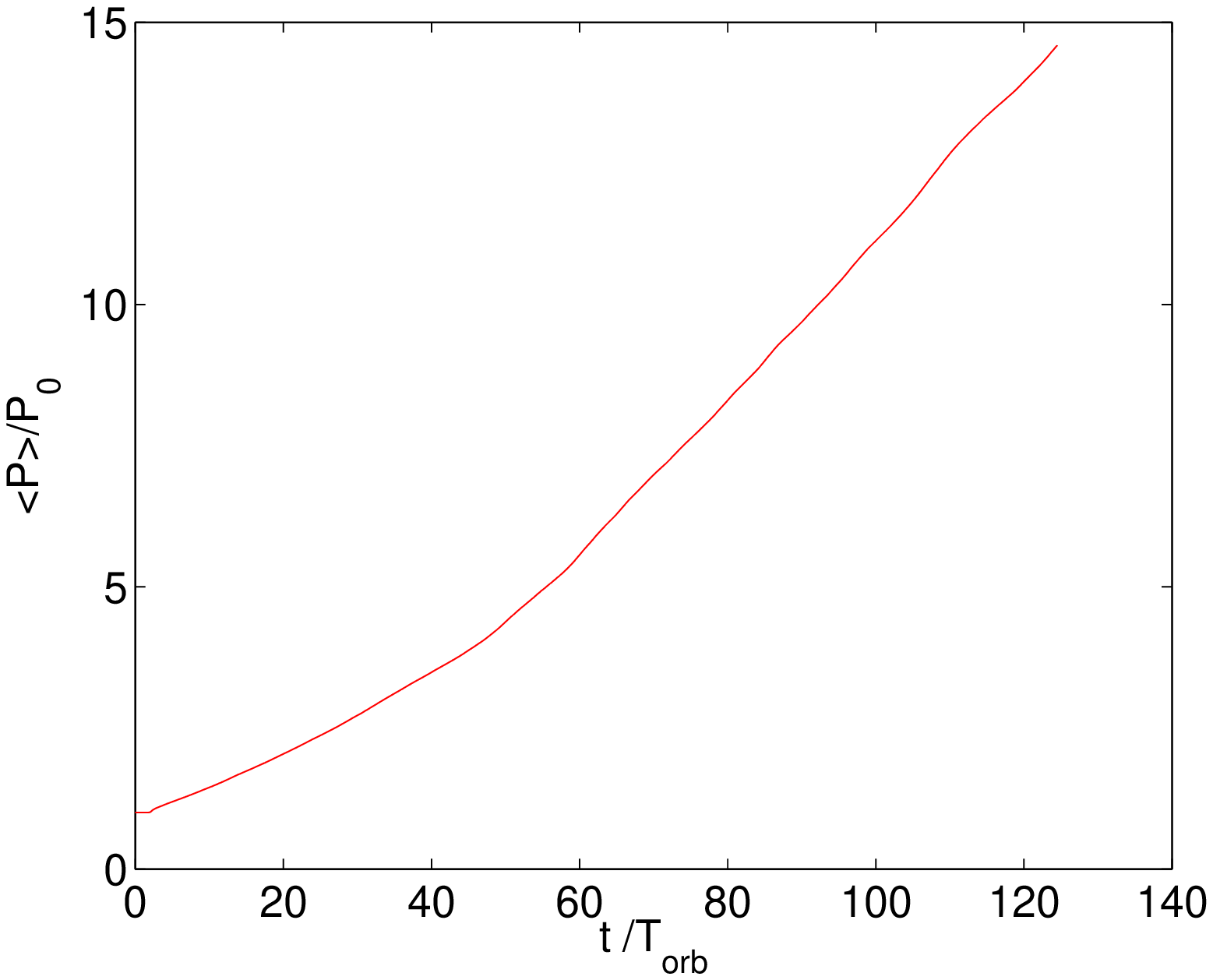}
\includegraphics[width=8cm]{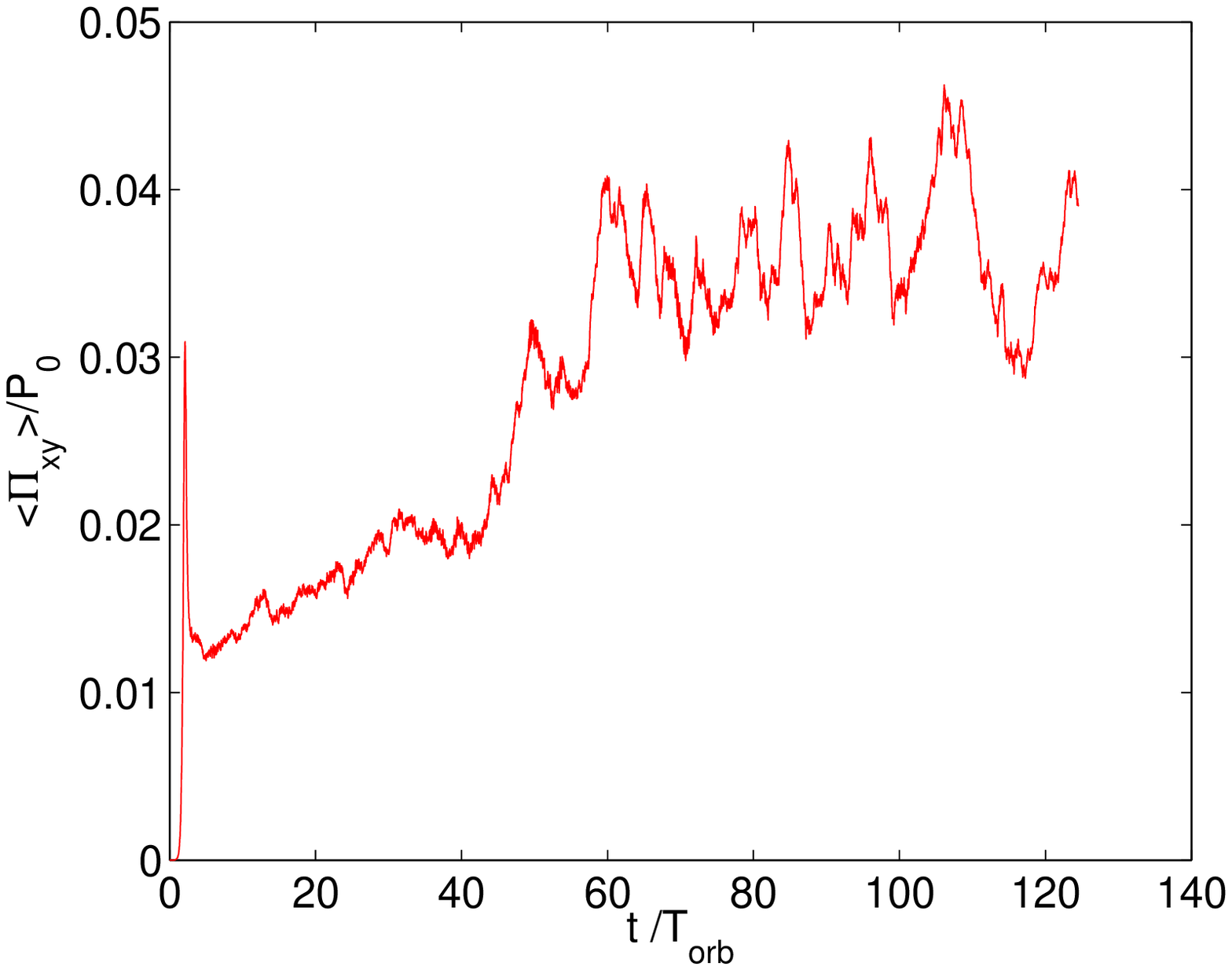}
\caption{ The time evolution of the gas pressure $P$ (left panel) and the total stress normalized by the
  initial gas pressure (right panel) for the zero-net-flux $L=4H_{0}$ simulation.}
\label{fig::znf4ST}
\end{figure*}

\section{Zero-net-flux simulations}

We cover the three possible topologies of the magnetic field in two
separate sections, starting with zero-net flux. First we present our
main results which display the effect of increased gas pressure on the
stress, and how the box size and the diffusion scales
impact on this behaviour. Cooling runs are
presented next, where a similar dependence is observed. Finally,
we look at isothermal simulations of differing sizes to see if these
trends are reproduced in simulations that are in quasi-equilibrium.
We place additional tests in Appendix B that reinforce our results.

\subsection{Heating runs}

\subsubsection{Influence of the box size}

We start off by considering boxes with no cooling ($\Lambda=0$) and no
explicit diffusion ($\nu=\eta=0$). The simulations, however, heat up
by numerical dissipation which is captured by our total energy
conserving scheme. We examine boxes with radial and vertical extents of $L=H_0$,
$2H_0$, and $4H_0$, all with $\Delta=H_0/64$. 

The time histories of the pressure and stress of the $L=4H_{0}$
simulation may be viewed in Fig.~\ref{fig::znf4ST}.
While the pressure shows the monotonic increase expected, the stress's evolution
is more complicated. First, it undergoes an exponential growth,
corresponding to the onset of the linear instability,
followed by a rapid decrease in stress as the initially ordered flow
breaks down into turbulence. This phase only takes a handful of
orbits. For the next 60 orbits, the stress exhibits significant
growth, of almost a factor of 3. During the same phase, the pressure
has increased by a factor of about 5. After 60 orbits the system
enters a third phase: the pressure carries on growing but the stress
plateaus and suffers large amplitude bursts (in contrast to the more
placid earlier stages). In summary, the system, adheres rather closely to the
expectations outlined in Section 2.2: after the linear phase, the box
heats up, with the stress following behind, but once the turbulent
eddy sizes hit the box size they can grow no more
and the stress reaches a constant level. 

\begin{figure}
\includegraphics[width=9cm]{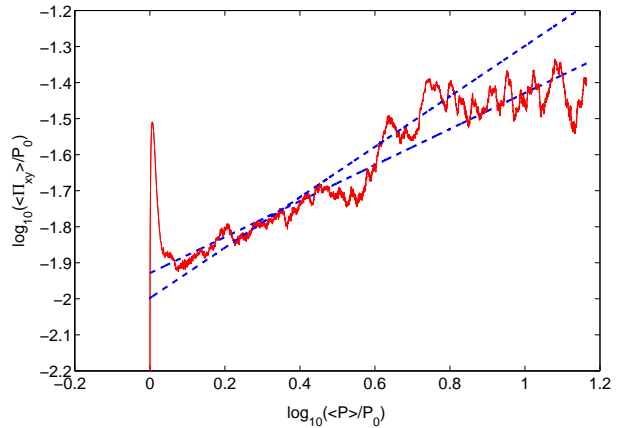}
\caption{The time evolution of the stress normalized by the
  initial gas pressure, as a function of the volume averaged pressure
    for a $L=4H_{0}$ simulation (red solid curve). The blue dashed lines
  have slopes $0.5$ and $0.7$.} 
\label{ZN4HStressvsP}
\end{figure}

In Fig.~\ref{ZN4HStressvsP} we plot the stress as a function of
pressure, in order to estimate $q$. As mentioned earlier, this is not
without ambiguity and we overplot lines of $q=0.5$ and $q=0.7$ to
indicate possible ranges for this quantity. Despite the uncertainty
it is clear that, before the box size interferes, there is a
relatively strong correlation between $P$ and $\Pi_{xy}$. 
This is in
marked contrast
to earlier work. 

\begin{figure}
\includegraphics[width=9cm]{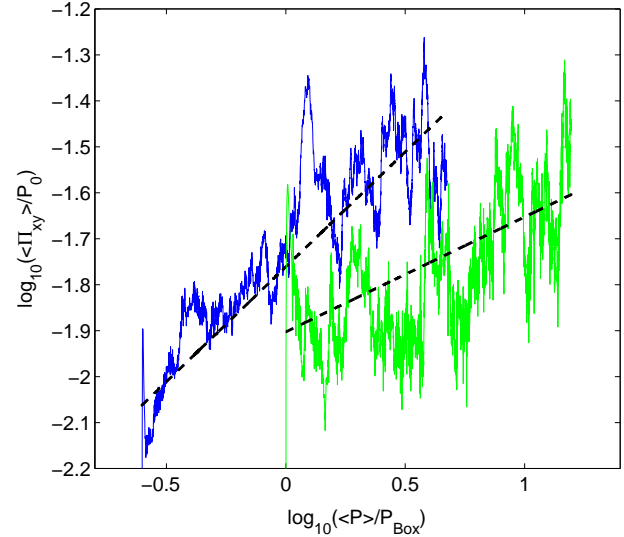}
\caption{The time evolution of the total stress
 against the gas pressure normalized by
 $P_{Box}=P_{0}L^{2}/H_{0}^{2}$. The blue curve is from the $L=2H_{0}$
 simulation and the green curve is from the $L=H_{0}$ simulation. 
Lines with slopes $0.25$ and $0.5$ are included for comparison.}
\label{fig::SPscaled}
\end{figure}

As argued in Section 2, we believe that previous simulations gave
lower $q$s as a
result of insufficiently large boxes. To test this idea, we ran
simulations with smaller $L$. The results are shown in 
Fig.~\ref{fig::SPscaled}.
For $L=H_0$, we obtain only a minimal
increase of stress with pressure. We find that $q=0.1 - 0.3$
which is consistent with the $q=0.25$ achieved by SITS04, who used
similarly sized boxes. The $q$ increases when we move to
$L=2H_{0}$. We find then that $q=0.4-0.6$. Though the
determination of $q$ is made difficult by the enhanced burstiness of the
signal, the result is clear: the $H_0$ box is too small
to adequately describe the growth of the stress. In addition, the fact that $q\approx 0.5$ in
both the $2H_0$ and $4H_0$ boxes suggests that $q$ has converged with
respect to $L$ in the $2H_0$ box. 

\subsubsection{Influence of the grid}

It must be emphasised that the result $q\approx 0.5$ is
still very much determined by the numerical parameters. On dimensional
grounds (cf.\ Section 2.1), any deviation from $q=1$ \emph{must} arise
from either a dependence on the box size, on the grid scale, or on
both. In the previous subsection we investigated the influence of $L$,
in this subsection we investigate $\Delta$.  

Fromang \& Papaloizou (2007) found that the stress is proportional to
the grid scale $\Delta$ in their isothermal simulations. Suppressing
the box size dependence for the moment, these results suggest the
scaling $\Pi_{xy} \sim \rho  
 \Delta\Omega c_s \sim P^{1/2}$ in agreement with the $q\approx 0.5$ we
find. To reproduce this scaling in our heating runs
we took the $L=2H_0$ box and tried
resolutions of $\Delta= H_{0}/16,\, H_{0}/32$, in addition to the fiducial
$H_0/64$. In all cases $q\approx 0.5$. In addition, during both the growth phase and
the plateau phase, the magnitude of the stresses is
proportional to $\Delta$,
with the
stress plateau taking values $\approx 0.12,\,0.063,$ and $0.03$
for the three $\Delta$s tried.

In summary, this sequence
suggests that the stress scales as
\begin{equation}  
\Pi_{xy}\sim \rho \Delta \Omega c_s,
\end{equation}
during the first stages of the simulation (when the influence of the
box size is mitigated). Afterwards it scales as
\begin{equation}
\Pi_{xy} \sim \rho \Delta L \Omega^2,
\end{equation}
during the plateau stage, once $H$ grows sufficiently large. 

Certainly, these results do not meet all the predictions of
  Sections 2.1 and 2.2. Though
$q$ increases in larger boxes, it does not approach the
value 1. Moreover, the maximum stresses achieved are
too small, proportional to $L$ not $L^2$. 
Obviously, the grid is
obstructing the growth of the stress, preventing it from (a) fully responding to
the pressure and (b) from obtaining larger values.
 As first shown by Fromang \& Papaloizou
(2007), MRI turbulence prefers to anchor 
itself on the grid scale, and
this imposes a constraint comparable to that enforced by the
acoustic radiation. In fact, if the fluid velocities follow $c_s$ but the turbulent lengthscales
are stuck on the grid, so that $l_\text{turb} \sim \Delta$, 
then we obtain the simulated result $q\approx 0.5$.  

\begin{figure}

\includegraphics[width=9cm]{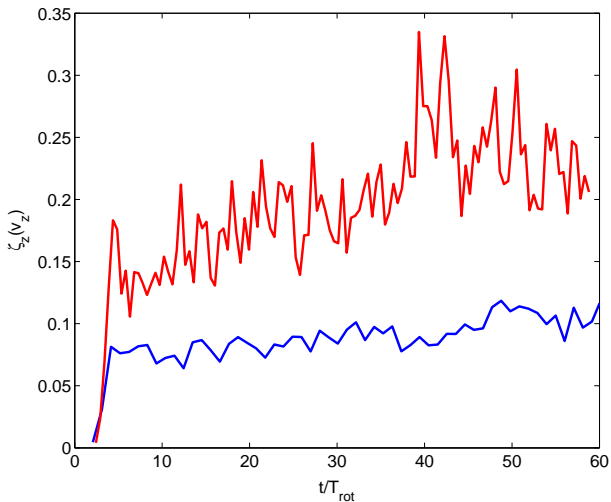}
\caption{The time evolution of the correlation length, defined in
  Eq.~\eqref{eq::cor}, for the $L=4H_{0}$ simulation without physical
  diffusivities (blue curve), and the $L=4H_0$ simulation with
  physical diffusivities (red curve).}
\label{fig::znf4cor}
\end{figure}

To test this idea
we calculate the
correlation length for the $4H_0$ box and plot the outcome
in Fig.~\ref{fig::znf4cor}. If $l_\text{turb}$ was stuck on the grid
then the correlation length would stay constant with time. Actually,
the figure reveals a modest
increase. 
Over 60 orbits this is some $\approx 1.4$, after
which $\zeta_{z}(v_{z})$
plateaus. Considering that the stress only grows by a factor 3, the
increase in $\zeta_{z}(v_{z})$ 
is not negligible and reveals that the correlation length is not
entirely anchored to $\Delta$, though it cannot wander too far away.

In conclusion,
these results indicate
that it may be impossible to obtain numerical convergence when
determining the stress-pressure relationship in zero-flux simulations
of this type. 
Though the
influence of the box size may be mitigated to some extent, the
influence of the grid size is fundamental and cannot be escaped.
The stress is proportional to $\Delta$, which forces $q=0.5$ on
dimensional grounds. This problem potentially limits the
application of the simulations to physical situations.

\subsubsection{Influence of physical diffusion}

In order to further probe what the small scales are doing, we
undertook simulations with explicit diffusion coefficients.
Fromang (2010) found that, in isothermal runs, the stress is approximately
independent of the Reynolds number for $P_m=4$ when
$Re$ takes values between $Re=3125$ and $12500$.
This suggests that the influence of the physical diffusion scales (as opposed
to the numerical diffusion scales) may disappear if they are forced to
be sufficiently
small. 

\begin{figure}
\includegraphics[width=8.5cm]{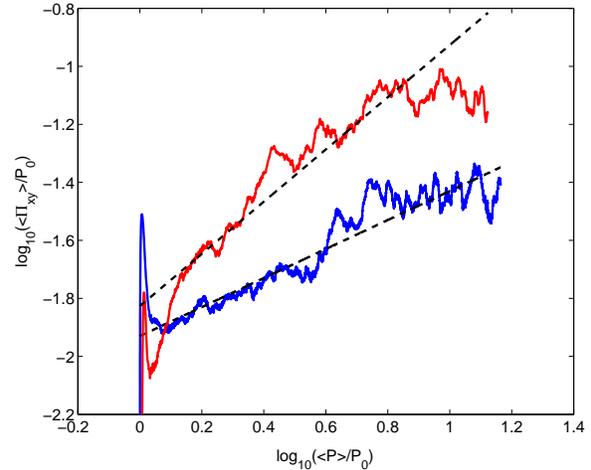}
\caption{The time evolution of the total stress normalized by the
  initial gas pressure, as a function of the volume averaged pressure
  for $L=4H_{0}$ simulations. The red curve is from a simulation with
  explicit diffusivities. The blue curve is from a simulation
  identical in all respects other than that explicit diffusivities are
  omitted. Lines of slopes $0.5$
 and $0.9$ have been superimposed for clarity.}
\label{fig::znfIC}
\end{figure}

To test this we adopt boxes of size $H_0$, $2H_0$ and $4H_0$ with
$Re=1250$
and $Rm=5000$. 
When $\Delta = H_0/64$, the resistive scale is resolved and the
viscous scale is marginally resolved (see Appendix B for more
details). As in the previous subsection, we find that when the box
size increases so to does $q$: from $\approx 0$, to 0.5, and finally to 0.9 in the largest box. 
The time evolution of stress against $P$ is plotted in
Fig.~\ref{fig::znfIC} (red curve) for the $4H_0$ box. For comparison we have overplotted
the curve of the diffusionless $4H_0$ run (blue curve). 

The most striking result, of course, is that in sufficiently
  large boxes $q$ can achieve a
  value close to 1, which is more in line with
our initial expectations. In addition, the stress increases to
larger values than earlier. Physical diffusion elicits dramatically different
behaviour vis-a-vis the numerical grid. 
It would appear that the former does not constrain the turbulent
eddies nearly so forcefully, leading to a stress that can grow more
freely. In Fig.~\ref{fig::znf4cor} the correlation length is plotted,
which further illustrates the contrast between the two cases. 

The reason for why $q$ is not exactly 1 may
be due to the residual influence of the box, or possibly the influence
of the Reynolds numbers. The fact that the viscous scales are only
marginally resolved could also play a role. But note 
 that the stress only increases by a
factor of some 5 before it plateaus, not the factor 16 we might expect
if $l_\text{turb}$ increased from $H_0$ to $4H_0$. 
The reason seems to be that the
plateau phase starts before $H$ reaches the box size, such that $H$
increases by a factor only $\sim 2.5$ during the growth phase of the
stress, and not the anticipated factor 4 (Fig.~6). 
Simulations with yet bigger boxes as well as higher Re
and Rm, though numerically expensive, could help better understand
what is going on here.
For the moment we limit ourselves to emphasising the
striking difference between the case with explicit diffusion and the
case without. The former exhibits a stress-pressure relationship much
closer to the standard alpha model, and moreover has the potential to
give converged answers with respect to the numerical parameters.

\subsection{Cooling runs} 

So far we have captured the stress-pressure relationship by
heating up the system (increasing $P$) and seeing what happens to the stress.
A complementary approach
is to observe the stresses as the system cools (i.e.\ as $P$
decreases). If there truly is a meaningful correlation between the
two, then the stress must increase and decrease at the same rate in
the two cases.

\begin{figure}
\includegraphics[width=9cm]{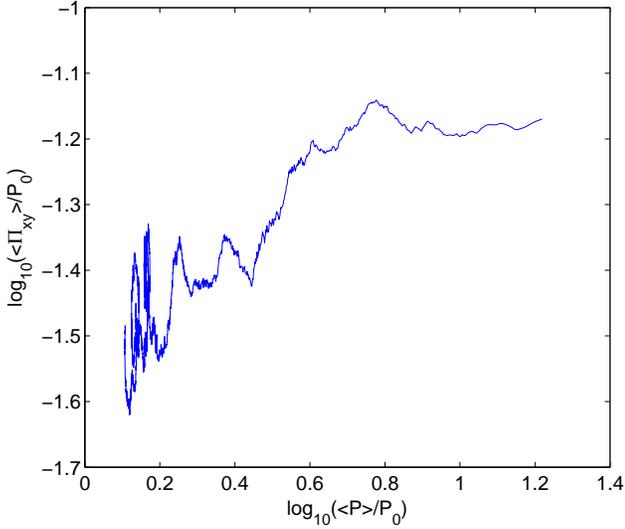}
\caption{The time evolution of the total stress normalized by the
  initial gas pressure, as a function of the volume averaged pressure
  for a $L=4H_{0}$ simulation (blue solid curve) when a constant
  cooling is introduced. Note that the system evolves from right to
  left. This should be compared to the analogous heating run in Fig.~3.}
\label{fig::cooling}
\end{figure}

We take a $L=4H_0$ box with lower resolution $\Delta=H_0/32$, no
physical diffusivities, and
initially impose no cooling. The fluid heats up 
until the thermal pressure increases
by over an order of magnitude and $H\approx L=4H_0$. 
At this point, we introduce a cooling law with $m=2$ and $\theta$
chosen to introduce a stable thermal fixed point at $P\sim P_{0}$,
i.e. $H=H_0$. The fluid is then attracted to this cooler state.
The ensuing cooling phase of the simulation is plotted in
Fig.~\ref{fig::cooling}.

Unlike the heating evolution, described in Fig.~\ref{ZN4HStressvsP},
the system evolves from the top right to the bottom left but in most
other respects shares the same shape and importantly the same $q$. 
Once the system cools to the point that $P\approx 5 P_0$, the
plateau stage ends, and the stress starts decreasing. The main difference
is that $\Pi_{xy}$ is systematically larger in the cooling run as
compared to the earlier heating run; but this is due entirely
to the lower resolution used (see previous subsection). Also at the
end of the cooling phase, when $H$ is small, the turbulence becomes
unusually bursty. This is due possibly to the increased proximity of the system
to criticality ($H$ is closer to $\Delta$).

\subsection{Isothermal runs}

In the previous subsections we described the stress-pressure
relationship 
in systems out of equilibrium. Following SITS04 and Pessah et
al.~(2007), 
we can also look at
isothermal systems 
that have reached a quasi-equilibrium and observe how the saturated
stress depends on the scales $L$, $H$ and $\Delta$. 
As emphasised earlier,
previous simulations use $L\lesssim
H$ and are in the box-dominated regime. We look at isothermal simulations in the
opposite limit, with a suite of simulations of fixed $H$ but of
$L=H,\, (3/2)H,\,2H,\,4H,$ and $8H$. 

First, we hold the resolution fixed per $L$, which means the resolution
per $H$ varies from run to run. This scenario mimics the heating runs
in Section 4.1 where, as the box heats up and $H$ grows, the number of
grid zones per $H$ increases. The $\alpha$s computed are listed in
Table I, where the rough scaling $\alpha \sim L/H$ is exhibited,
significantly weaker than Pessah et al.~(2007), who find
$\alpha\sim (L/H)^{5/3}$. 
This we attribute to
the fact that $L>H$.

\begin{table}
	\centering
	    \begin{tabular}{| l | l | l | l |}
		    \hline
		    L/H & $\Delta/H$ & $ \alpha $ & Type\\ \hline \hline
		    1 & 1/64 & 0.0110  & ZNF \\ \hline
		    3/2 & 3/128 & 0.0159 & ZNF \\ \hline
		    2 & 1/32 & 0.0199 & ZNF \\ \hline\hline
		    1 & 1/32 &  0.0137 & ZNF \\ \hline
		    2 & 1/32 & 0.0199 & ZNF \\ \hline
		    4 & 1/32 & 0.0246 & ZNF \\ \hline
		    8 & 1/32 & 0.0301 & ZNF \\ \hline\hline
		    1 & 1/32 & 0.0154 & Tor \\ \hline
		    2 & 1/32 & 0.0214 & Tor \\ \hline
		    4 & 1/32 & 0.0287 & Tor \\
		    \hline
	    \end{tabular}
	    \caption{The average $\alpha$ for isothermal
              simulations with varying $L$ and $\Delta$. The two field
              configurations are zero-net-vertical flux (`ZNF') and net
              toroidal flux (`Tor'), the latter employs $\beta=1000$.}
	    \label{tab::isores}
\end{table}

\begin{figure}
\includegraphics[width=8cm]{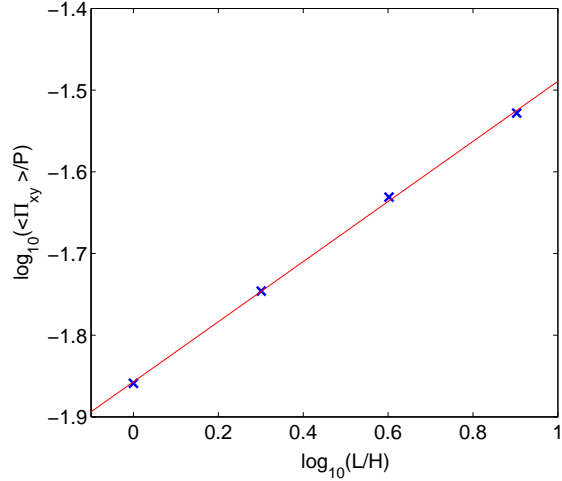}
\caption{Total stress, normalised by the pressure, as a function of
  the ratio of the box size to scale height for isothermal
  zero-net-flux simulations. For comparison, a curve with $\alpha
  \propto (L/H_{0})^{0.37}$ is overlaid. }
\label{fig::iso}
\end{figure}

The turbulent $\alpha$ must also be proportional to a
dimensionless function of the ratio $L/\Delta$. In order to determine this function,
we undertake a set of simulations exploring a larger range of
box sizes and with the number of
grid points per $H$ remaining constant.
The calculated $\alpha$s are listed in Table \ref{tab::isores} and plotted as a
function of $L/H$ in Fig.~\ref{fig::iso}. Now we find the rough
scaling $\alpha \sim (L/H)^{2/5}$. Combining the two scalings from the
two sequences of runs
yields an expression for the stress
\begin{equation}\label{isos}
\Pi_{xy} \sim \rho H \Delta^{3/5} L^{2/5}\Omega^2,
\end{equation}
which holds for isothermal simulations in which $H<L$. 

Estimate \eqref{isos} raises a number of points. First,
it implies that  $\Pi_{xy}\sim
P^{1/2}$, yielding a $q$ in agreement with our heating runs of Section
4.1, a cross-validation that inspires confidence in both results.
Second, the stress is $\sim\Delta^{3/5}$, in contrast to
 Fromang \& Papaloizou (2007) and Pessah et al.~(2007), who find it
 proportional to $\Delta$. This puzzling disagreement could be due to
 the fact that our simulations are in the $L>H$ regime, or it could
 perhaps issue from different box aspect ratios. Certainly the
 discrepancy encourages further work.
Third, the system 
cannot escape the influence of the box size, even in the largest of the
simulations. We expected that when $L=8H$ there would be
a separation of scales between $L$ and $l_\text{turb}$
(whether the latter is set by $H$ or $\Delta$) and that the turbulent
eddies might no longer feel $L$. This is evidently not the case. 
It suggests that
large-scale structures, magnetic or acoustic, play some role even in
zero-net-flux turbulence (see also Guan et al.~2009, hereafter
GGSG09). These structures can develop
because the simulation has been run for a long time (in order to reach
equilibrium) and the
box has begun to `sense' its finite size and its
periodicity. This is probably not the case in the heating runs of
Section 4.1, and one might conclude that isothermal (or
any long-time equilibrium) simulations are inappropriate for
determining the instantaneous stress-pressure relationship.

\section{Net-flux simulations} 

\subsection{Toroidal fields}

We now examine the effect of the vertical box size on the stress-pressure
relationship in net-toroidal field
simulations. Explicit diffusion coefficients are omitted, but the
boxes
 are permitted to heat up via numerical dissipation. We use,
 $\Delta=H_0/32$ and set $L=H_0,\,2H_0,$ and $4H_0$. The strength
 of the net field is fixed by the initial beta, $\beta= 200$. The
 corresponding Alfv\'en length is then $\l_{A}=\sqrt{2/\beta}H_0$, which is
 roughly 3 times $\Delta$. Thus the input scale of the
MRI turbulence is resolved, but there is no inertial range to speak
of --- a serious deficiency of our, and most extant, MRI simulations
(but see Fromang 2010 and Meheut et al.~2015). 

Turbulence in net-toroidal simulations develops slowly, over
tens of orbits, irrespective of the 
temperature increase. This initial phase complicates the interpretation of the
stress's growth and makes attributing a well-defined $q$ problematic.
In order to oversome this obstacle,
we remove the slow non-linear development of the turbulence. 
Our strategy is to introduce cooling at the beginning of the run and
let the box come to thermal equilibrium with a quasi-steady $H$ near
a target $H_0$. The cooling function adopted is $\Lambda=\theta P^{2}$. 
Once the system has remained in this quasi thermal equilibrium for
$\sim 30$ orbits the cooling is removed 
and the pressure allowed to increase. Subsequently, any increase in stress will
be due to the change in pressure alone and not due to the initial long transient.

\begin{figure}
\includegraphics[width=8cm]{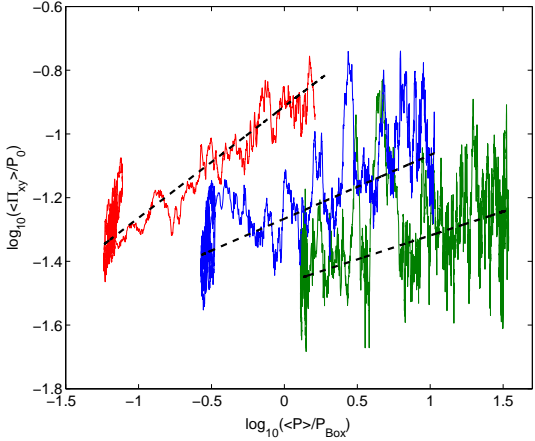}
\caption{The time evolution of the total stress normalized by the
  initial gas pressure, as a function of the volume averaged pressure
  for the net-toroidal simulations with $L=4H_{0}$ (red curve),
  $L=2H_{0}$ (blue curve) and $L=H_{0}$ (green curve). The superimposed
   lines have slopes $0.15, 0.2, 0.35$.}
\label{fig::TorPre}
\end{figure}

Results from these simulations are shown in
Fig.~\ref{fig::TorPre}. We find that $q$ increases with box size, from $\approx 0.15$
in the $L=H_{0}$ simulation to $\approx 0.35$ in the $L=4H_{0}$.
The stress-pressure relationship is weaker than for the zero-net-flux
case, but we can still discern the influence of the box size in
limiting $q$. It is unclear, however, if $q$ has converged yet with
respect to $L$.

\subsubsection{Field strength dependence}

The stress in net-flux simulations
is not only governed by $c_{s}$ but also by the strength of the
imposed magnetic field. Indeed HGB95 found that
$\Pi_{xy}\propto E_B \propto v_A$
in their early net-vertical and net-toroidal flux simulations, where $E_B$
is the total magnetic energy. The same scaling was noted by GGSJ09, who
argued that $\Pi_{xy} \propto c_s v_A$. Assuming this holds for thermally
evolving systems, this means the stress must be proportional to the square
root of $P$, i.e.\ $q = 0.5$. The value of $q$ measured in the last subsection is
marginally consistent but slightly smaller than this scaling. The
difference may be attributed to the remaining influence of the box
size, the grid, or the strength of the imposed field, which we explore now.

To study the effect of the imposed flux
 we perform $L=4H_{0}$
simulations with $\Delta=H_{0}/32$ but with $\beta=50,\, 2\times 10^{2},\,
5\times 10^{2}$, $5\times 10^3$, and $10^{4}$. We first let the
simulations settle into a thermal equilibrium before setting
$\Lambda=0$ and letting them heat up.

\begin{table}
	\centering
	    \begin{tabular}{| l | l | l | l |}
		    \hline

		    $\beta$ &  $v_{A}$ & $ \langle\Pi_{xy}\rangle/P_{0}$ & $q$ \\ \hline \hline
		    50 & 0.200 & 0.098  & 0.1-0.2 \\ \hline
		    200 & 0.100 & 0.051 &  0.3-0.4 \\ \hline
		    500 & 0.063 & 0.033 & 0.4-0.5  \\ \hline
		    5000 & 0.020 & 0.024 & 0.4-0.5 \\ \hline
		    $10^{4}$ & 0.014 & 0.025 & 0.4-0.5\\
		    \hline
	    \end{tabular}
	    \caption{Average total stresses calculated in the initial thermal equilibrium of
             the net toroidal simulations described in Section 5.1.1. as well as estimated $q$
             values using these states as restarts.}
	    \label{tab::Toreq}
\end{table}

Firstly, during the
thermal equilibrium stage when $P$ is quasi-steady, we find that
$\Pi_{xy}\propto v_{A}$ if $l_{A}>\Delta$, Table
\ref{tab::Toreq}. Thus our diabatic equilibrium simulations agree with previous
isothermal runs (GGSJ09) and heating runs (HGB95). Note that when the magnetic tension is not
resolved, $l_A<\Delta$, the average stress becomes independent of field strength.
            
\begin{figure}
\centering
\includegraphics[width=8cm]{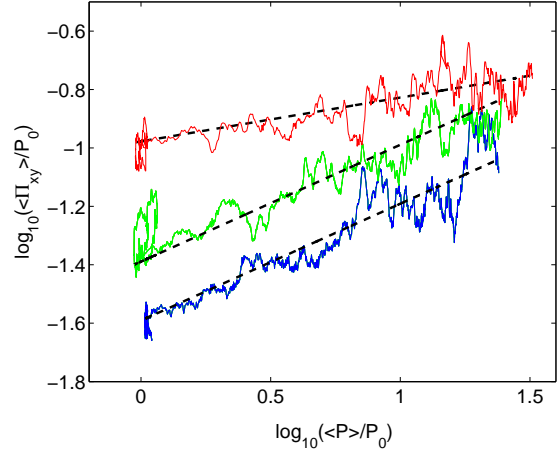}
 \caption{Stress against pressure for $L_{z}=4H_{0}$ net-toroidal
   simulations with $\beta= 50,\, 2\times 10^{2}$ and
   $10^4$ (red, green, and blue respectively). 
 Dashed lines with slopes $0.15$, $0.35$ and $0.4$ have been superimposed (in descending order).} 
             \label{fig:torEq}%
\end{figure}

  \begin{figure}
\centering
\includegraphics[width=8cm]{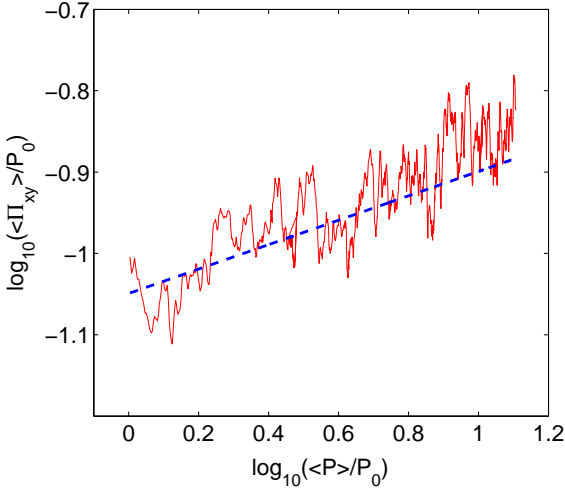}
 \caption{Stress against pressure for the $L=2H_{0}$ net-toroidal
   simulations with $\beta= 50$ and reduced heating. The dashed line has slope $0.15$.} 
             \label{fig:torEqr}%
\end{figure}

Secondly, during the heating phase of the simulations, we find that
the stress-pressure relationship also depends on $v_A$. 
Our results are shown in Fig.~\ref{fig:torEq}, which reveal that 
$q$ is an increasing function of $\beta$. For our moderate strength
toroidal fields simulations, $\beta=200, 500$, we obtain $q\approx
0.35$. 
Increasing the field strength decreases $q$ to $~0.15-0.2$, while for our weak field simulations $q=0.35-0.5$. 

The weak field runs are easy to understand: the Alfv\'{e}n length, $\l_{A}$, 
is less than the grid $\Delta$. Consequently, the fluid barely `feels' 
the magnetic tension from the imposed field and instead behaves as if
the box were zero-net flux. This explains the larger $q$ for the $\beta=2\times
10^{2}$ and $10^{4}$ simulations, which approach the results of
Section 4.1.

When $l_A$ is resolved, 
the reason for the weaker dependence of stress on pressure is
more difficult to attribute. One idea is the following: 
low $\beta$ simulations
exhibit rapid heating, the speed of which could outstrip the
ability of the fluid to adjust to the rising temperature. As a
result, the stress-pressure relationship could be weakened. To test
this 
we slow down the heating via a cooling law that is precisely half the heating
rate, i.e.\ $\Lambda=\frac{1}{2} dP/dt$. The result is plotted
in Fig.~\ref{fig:torEqr}, where we see no change in the measured
$q$. Thus very fast heating is not the culprit in the low-$\beta$
low-$q$ connection. 

Could the weak dependence be a result of a too-small azimuthal box
size? GGSG09 show that the longest horizontal correlation
lengths are less than $H$ in isothermal simulations with $\beta=100$
and $400$. This suggests that even at $\beta=50$ the turbulence is 
unrestrained by the domain's azimuthal size. To check, however, we
undertook a simulation with $L_x=L_z=4H_0$, $L_y=10H_0$, and
$\beta=50$. The measured $q$ is little different to the smaller box
 with the maximum horizontal correlation length $\lesssim
H$.

The most likely explanation, perhaps, is that a strong imposed magnetic field   
interferes with the manner in which acoustic radiation
limits the flow, possibly by altering the nature of the pressure waves
or by directly impeding the turbulent motions themselves.
 Magnetic tension may not only enable the MRI
 but also restrict its nonlinear
development, especially on smaller
scales. Dedicated simulations could help test and further develop this
idea.

\subsubsection{Isothermal runs}

For completeness we also performed a number of isothermal simulations,
identical to the second set of simulations of Section 4.3 (where the
number of grid points per $H$ is kept constant) but with net toroidal
magnetic flux and $\beta=1000$. The simulations
are summarised in Table I. 

\begin{figure}
\includegraphics[width=8cm]{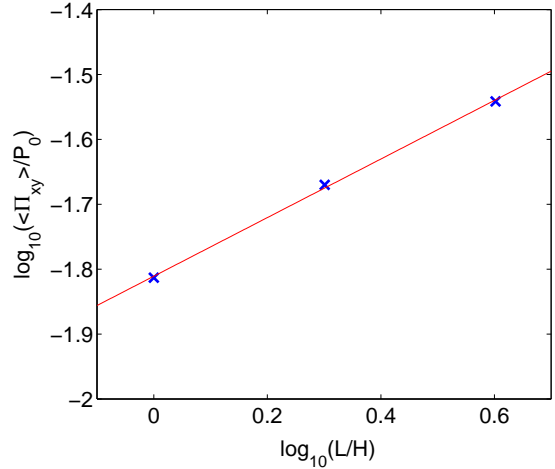}
\caption{Total stress, normalised by the pressure, as a function of
  the ratio of the box size to scale height for isothermal
  net-toroidal simulations. For comparison, a curve with $\alpha
  \propto (L/H_{0})^{0.45}$ is overlaid.}
\label{fig::isotor}
\end{figure}

This sequence yields approximately 
the same relation between saturated stress and
box size as in zero-net flux 
simulations, $\Pi_{xy}\sim (L/H)^{2/5}$, plotted in
Fig.~\ref{fig::isotor}. 
On dimensional grounds, $\Pi_{xy}$ must also be proportional to a 
function of both $\beta$ and $H/\Delta$. Its $\beta$ dependence can 
be constrained by two GGSJ09 simulations at $\beta=100$ and 400, 
which show that the stress is $\sim \beta^{-1/2}$. 
For this range of magnetic field strength we then have
\begin{equation} \label{isotor}
\Pi_{xy} \sim \rho v_A\,L^{2/5} H^{3/5}\Omega\,f(H/\Delta),
\end{equation}
 where $f$ is a dimensionless function. 
GGSJ09 also conduct a resolution study and argue that once $l_A$ is
resolved the magnetic energy (and consequently, stress) becomes
independent of resolution. If true, then $f$ is approximately a
constant and Eq.~\eqref{isotor} yields $q=0.3$, in reasonable agreement
with the heating runs of Section 5.1, for similar $\beta$. 

Obviously, 
when the magnetic field is stronger
the scaling breaks down, and the $H$ dependence must
diminish. On the other hand, for weaker fields (when $l_A$ approaches
and then slips below $\Delta$) the $v_A$ dependence weakens. 
 A more comprehensive set of simulations probing a wider range of 
field strengths and resolutions may better constrain 
the behaviour of $\Pi_{xy}$.

\subsection{Vertical fields}

\begin{figure*}
\scalebox{0.5}{ \includegraphics{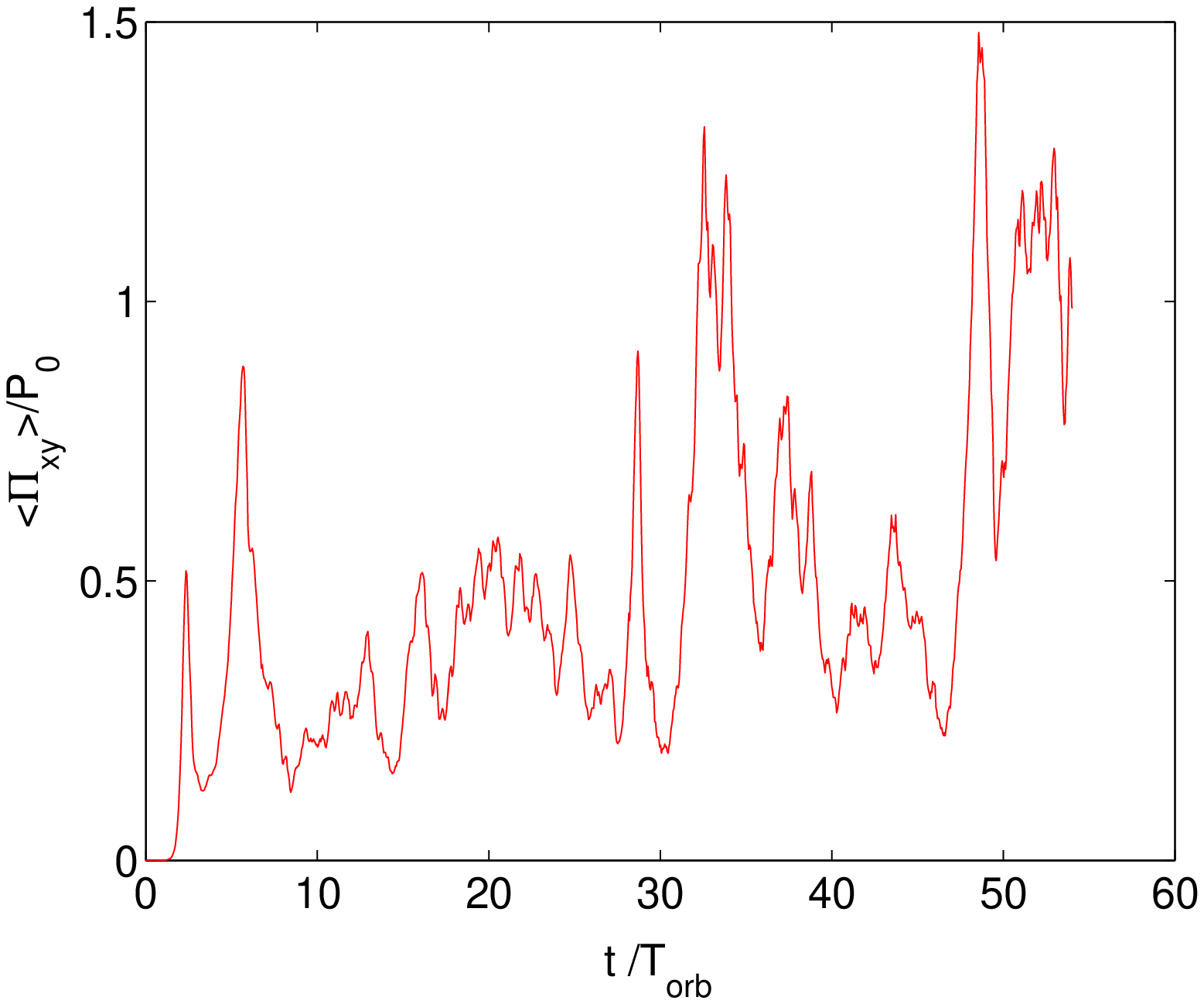}}
\scalebox{0.5}{ \includegraphics{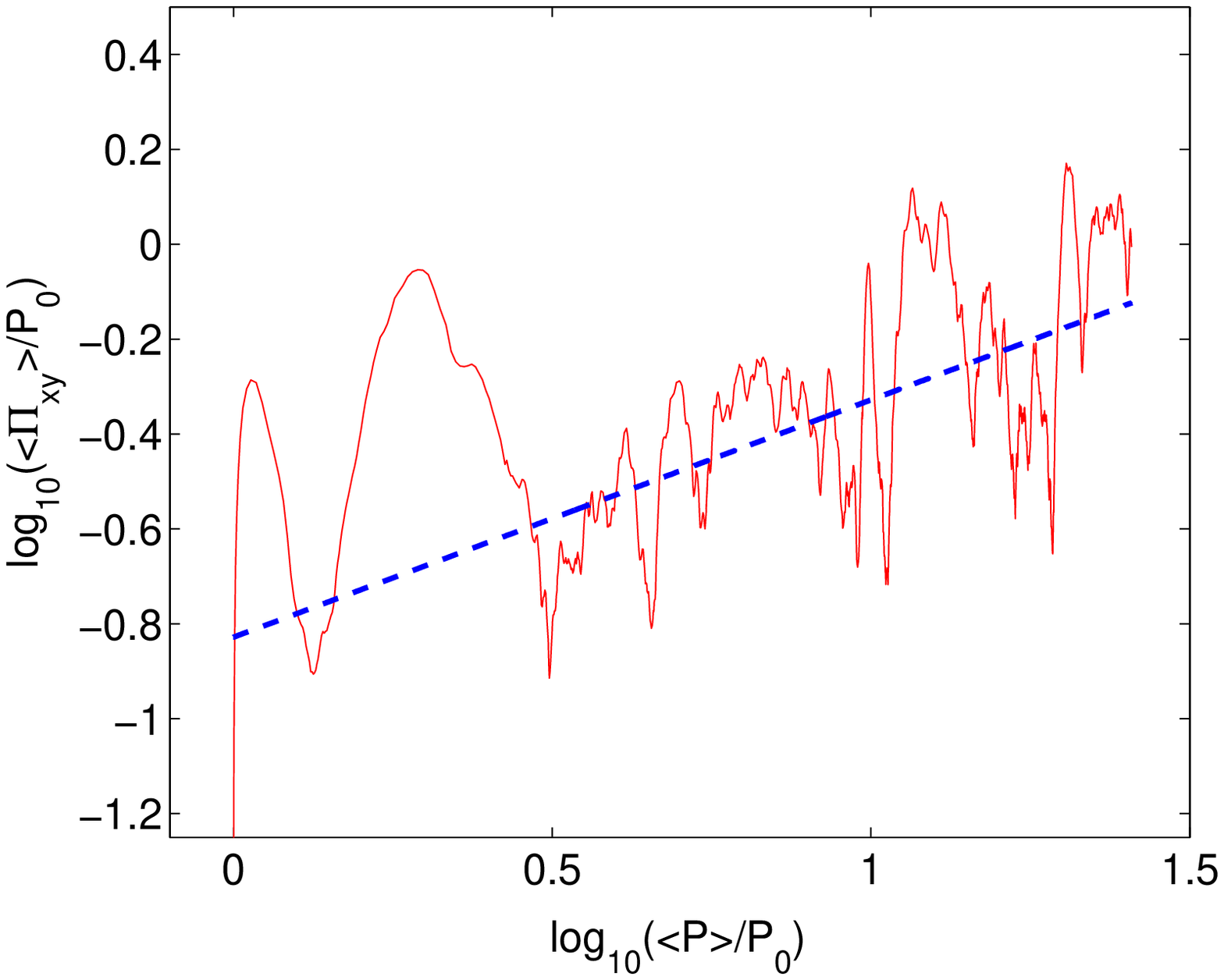}}
\caption{The time evolution of the total stress normalized by the
  initial gas pressure for the $(4H_{0},5H_{0}, 2H_{0})$ simulation
  with resolution $\Delta=1/32$, $\beta=500$ and reduced heating. The
  left panel shows $\Pi_{xy}$ as a function of time, the right panel
  as a function of $P$. The dashed line has $q=0.5$.}
\label{fig::vert24d}
\end{figure*}
Finally, we undertake a set of simulations with a net-vertical-flux
penetrating the computational domain. Our aim is to augment the
main trends of previous sections, rather than to be comprehensive.
The main result here is that $\Pi_{xy} \sim P^{1/2}$ in boxes of larger vertical
extent, and for weaker magnetic fields.

Simulation domains with a greater than unity aspect ratio, the ratio
of the radial to vertical lengths, exhibit
diminished channel modes (Bodo et al 2008). As channel bursts
distort the stress-pressure relation, we always use an appropriate
aspect ratio $L_x/L_z=2$ to minimize their influence. 
We first take a box of size
$(2H_{0}, 5H_{0}, H_{0})$ with $\Delta=H_{0}/64$ and set $\beta=1000$. 
The resulting relation that we obtain is very weak, 
$q=0-0.15$. We believe this to be in agreement with SITS04, who find
$q=0.17$, HGB95 and SHB09 who find no relation between
stress and pressure, 
and MHS15 who find a weak correlation or no relation at all depending on the numerical scheme. 

\begin{figure}
\includegraphics[width=8cm]{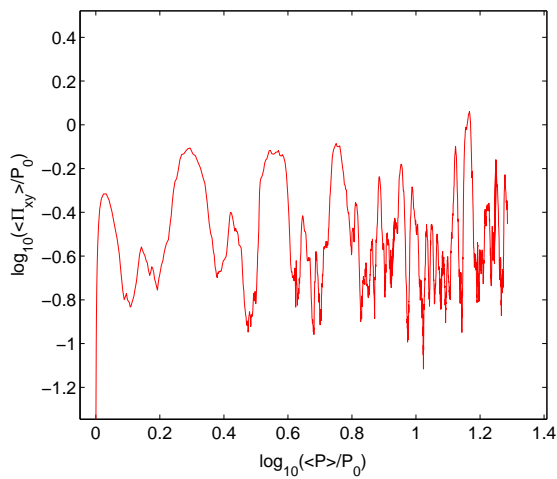}
\caption{The time evolution of the total stress normalized by the
  initial gas pressure, as a function of the volume averaged pressure
  for the $(2H_{0},5H_{0}, H_{0})$ simulation with $\Delta=1/32$,
  $\beta=500$ and reduced heating. The plot should be compared with
  the right panel of Fig.~\ref{fig::vert24d}.}
\label{fig::vert12rh}
\end{figure}

Increasing the box size to $(4H_{0}, 5H_{0}, 2H_{0})$ leads to
extremely rapid heating and the scale height exceeds the box size
within a few orbits. Disentangling initial growth from the pressure
dependence is
 impossible and therefore we introduce a cooling function, as in
 Section 5.1.1, to slow the heating: $\Lambda$ is set to $1/2$ the
 heating rate. We also reduce the resolution to $H_{0}/32$ and, to compensate, 
we increase the field strength so that $\beta=500$. The evolution of the
simulation is shown in Fig.~\ref{fig::vert24d}. After two initial
channel spikes there is an increase in stress with pressure over the
next $50$ orbits with approximately $q\approx 0.5$, a remarkable
contrast to the smaller box. For direct comparison, 
we plot, in Fig. \ref{fig::vert12rh}, log $P$ against
log $\Pi_{xy}$ from a smaller simulation of $(2H_{0}, 5H_{0}, H_{0})$ with identical cooling, resolution 
and $\beta$. Again, there is little to no correlation between stress
and pressure.

In principle, one could explore the connection between $q$ and
$\beta$. But decreasing $\beta$ further leads to more prominent
channel activity because the system is closer to criticality; these
bursts
 complicate the heating and stress behaviour and hence the extraction
 of a reliable $q$. Lower $\beta\,$s require larger $L_x$ and $L_y$,
 and more computationally expensive runs as a result. We leave open
 the question of the $\beta$ dependence of $q$ to future work, but
 expect similar outcomes to Section 5.1.

\section{Discussion and Conclusion}

The main aim of our paper is to highlight the importance of box size
in the relationship between MRI-induced stress and pressure. 
In most previous simulations the pressure scale height $H$ was greater 
than the vertical size of the computational domain $L$. Consequently,
turbulent eddies were  not restricted by the compressibility effects  
assumed by the alpha model, and so the associated stresses were unable to manifest a
meaningful dependence on pressure. 

In order to bring out the true
relationship between $\Pi_{xy}$ and $P$
we present simulations in the
opposite regime, when $H<L$. Generally, we find a stronger dependence,
and in zero-net flux simulations with explicit diffusivities we
almost reproduce $\Pi_{xy}\propto P$, as required by the $\alpha$-disk
model. The same simulations without explicit diffusivities, however,
cannot escape the influence of the grid and remain unconverged
with respect to the numerical parameters.

Toroidal-net-flux simulations witness a weakening of 
the stress-pressure relationship once the imposed magnetic field
gets too large. It is possible that this is to do with enhanced
magnetic tension, and consequently a 
greater proximity of the MRI to criticality, though how this works out
physically is unclear. Weaker fields in both net-toroidal and vertical
runs ($\beta >100$) roughly suggest the scaling
$\Pi_{xy} \lesssim c_s\,v_A$, which is consistent with GGSJ09, though
its origin is also mysterious and the role of numerical factors is still
to be fully determined.  

Our results present various puzzles and
problems that future simulations should pursue. Further work is required
to understand the zero-net-flux simulations with explicit
diffusivities. In particular, why does the growth in $\Pi_{xy}$ halt
`prematurely'? Simulations in large boxes and different Reynolds
numbers may help probe this behaviour. Other angles to take include
potential Prandtl number or box aspect ratio dependencies. 
Our toroidal flux simulations omit explicit diffusivities
but it would be beneficial to see what changes occur, especially in
$q$, when they are present. In the strong field, low $\beta$, 
regime the adjustment should be marginal. For weaker fields, 
when the Alfv\'en length is small, we anticipate more noticeable
discrepancies.
The $q$-$\beta$ connection in vertical-flux simulations
could also be explored more fully. Finally, it would be beneficial to extend 
our isothermal simulations. Of most interest would be
zero-flux runs with explicit diffusion, and toroidal-flux runs exploring
different $\beta$. 

An obvious generalisation of this work would be to include
vertical stratification. Unstratified
boxes can cleanly test the most fundamental idea of 
the alpha model: that acoustic radiation limits the development
of disk turbulence. However,
when $L>H$, the vertical structure of the disk should really be
included. Pressure controls the disk thickness and this geometric
effect (omitted in unstratified boxes) presents a second way that pressure
may influence $\Pi_{xy}$. Future work in this direction is challenging. 
In zero-net-flux simulations it may be difficult to escape   
the influence of the grid, due to resolution constraints. Net-vertical
flux simulations, on the other hand, may be complicated by the emergence of outflows
(Fromang et al.~2013, Bai \& Stone 2013, Lesur et al.~2013).

We finish with a short discussion on thermal instability in accretion
disks, the main impetus for this work. Essentially, thermal
instability must rely on the competing dependencies of the
heating and cooling on temperature. To establish
stability or instability we then must have knowledge of how $\Pi_{xy}$
depends on pressure. The classical instability of radiation-pressure
dominated accretion flows assumes that the stress is proportional to
total pressure (Lightman \& Eardley 1974, Shakura \& Sunyaev 1976, Piran
1978). Of course, radiation pressure is omitted in our simulations, but
a useful first step is to establish how $\Pi_{xy}$ depends on gas pressure
alone and to highlight constraints on the stress's evolution. This may
then aid in the interpretation of more advanced simulations,
especially as they appear to produce contradictory stability behaviour
(Hirose et al.~2009, Jiang \& Stone 2013). In particular, divergent
results are obtained in boxes of different $L_x$. 
A natural question is: does the stability's box-size
dependence issue from the kind of variable stress-pressure
relationship explored in our paper? Another question is: how captive
are these stability results to the numerical parameters, in particular
the grid scale $\Delta$? As shown in zero-net-flux runs, the stress is
propotional to $\Delta$ hence weakening its dependence on $P$ and
denying it numerical convergence. In the light of that, how are we to
interpret these simulations and then apply them to real systems?

\section*{Acknowledgments}
We thank the reviewer, Matthew Kunz, for a prompt and helpful set of comments.
We are also grateful to John Papaloizou, Gordon Ogilvie, Sebastien Fromang, Geoffroy Lesur and
Colin McNally for insightful discussions. Some of the simulations were
run on the DiRAC Complexity system, operated by the University of
Leicester IT Services,
 which forms part of the STFC DiRAC HPC Facility (www.dirac.ac.uk ). 
This equipment is funded by BIS National E-Infrastructure capital
grant ST/K000373/1 and STFC DiRAC Operations grant
ST/K0003259/1. DiRAC is part of the UK National E-Infrastructure.
run. JR and HNL are partially funded by STFC grants ST/L000636/1 and ST/K501906/1.
JG acknowledges support from the Max-Planck-Princeton Center for Plasma Physics.

\begin{appendix}

\section{Numerical tests}
In this appendix, we describe several numerical tests that have been performed to check the implementation in the code RAMSES of the shearing box source terms in the energy equation. A satisfactory second order convergence was obtained in all of these tests.

\subsection{Shearing waves}
In order to test the implementation of azimuthal advection, we have
performed numerical simulations of two types of particularly simple
shearing waves. The first exhibits an azimuthally varying entropy and density but uniform pressure:
\begin{eqnarray}
\rho &=& \rho_0\left\lbrack 1 + A\cos \left(\frac{2\pi n_y y}{L_y}\right) \right\rbrack, \\
P &=& P_0, 
\end{eqnarray}
where $\rho_0$, and $P_0$ are the background density and pressure, $A$ is a dimensionless amplitude of the entropy wave, $n_y$ the number of wavelengths in the azimuthal size of the box $L_y$.
The second possesses an azimuthally varying vertical magnetic field strength, but uniform entropy and total pressure (including magnetic pressure):
\begin{eqnarray}
\rho &=& \rho_0\left\lbrack 1 + A\cos \left(\frac{2\pi n_y y}{L_y}\right) \right\rbrack, \\
B_z^2 &=& B_0^2 - 2P_0\left\lbrack \left(\frac{\rho}{\rho_0}\right)^\gamma-1\right\rbrack, \\
c_s^2 &=& c_{s0}^2\left(\frac{\rho}{\rho_0}\right)^{\gamma-1},
\end{eqnarray}
where $B_0$ is the background vertical magnetic field strength, $c_{s}$ is the sound speed, $c_{s0}$ its background value.

These two types of waves are simply advected by the flow, and therefore sheared, without inducing any movement or pressure perturbation. The analytical solution of the time-evolution of the these waves is particularly simple
\begin{equation}
f(x,y,t) = f(x,y+xSt,0)
\end{equation}
where $f$ is any physical quantity, $S$ is the shearing rate. 

\begin{figure}
\centering
\scalebox{0.8}{ \includegraphics{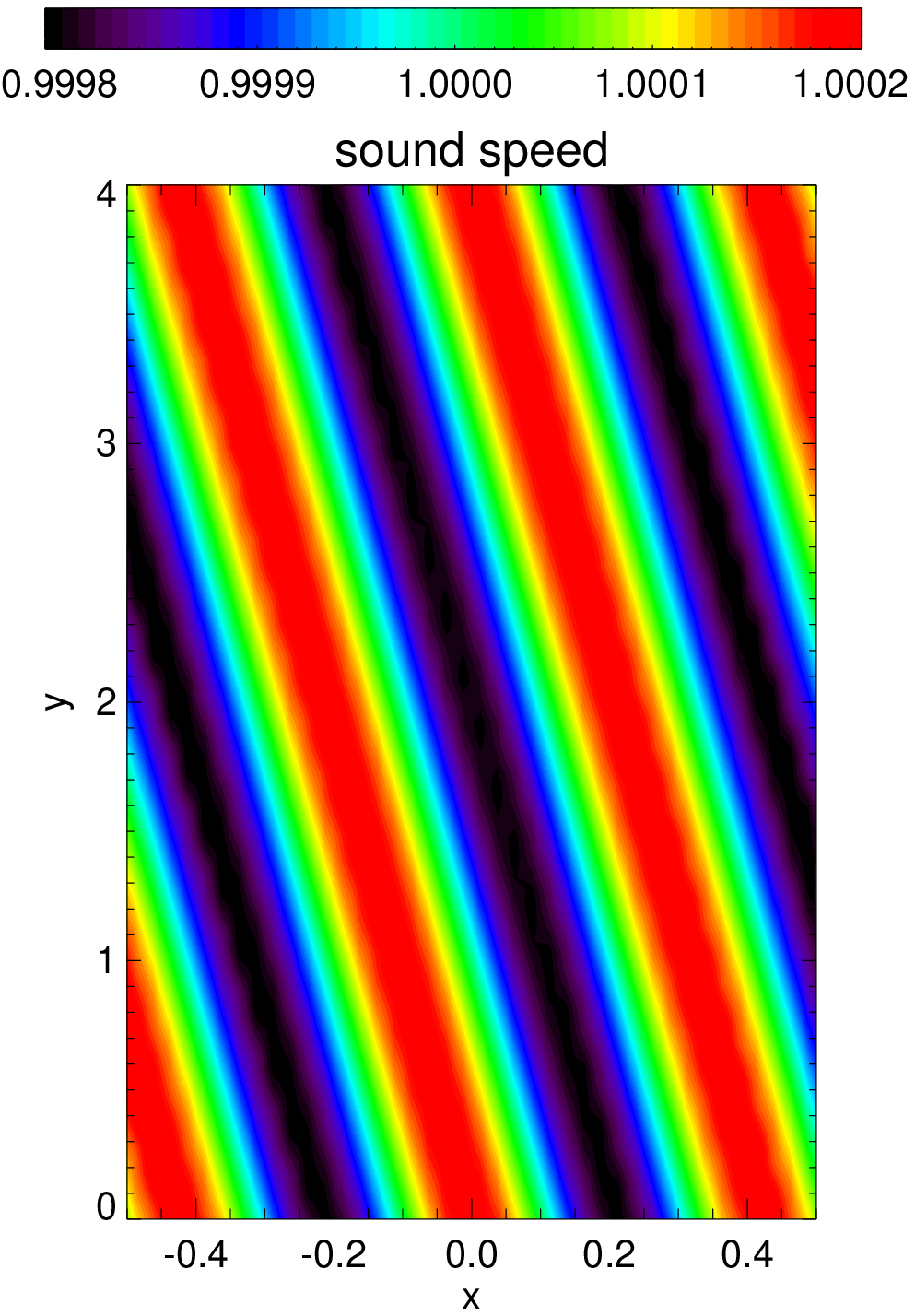}}
 \caption{Sound speed distribution after one orbit of evolution for the magnetic shearing wave.}
             \label{fig:test_shear_image}%
\end{figure}

\begin{figure}
\centering
\scalebox{0.8}{ \includegraphics{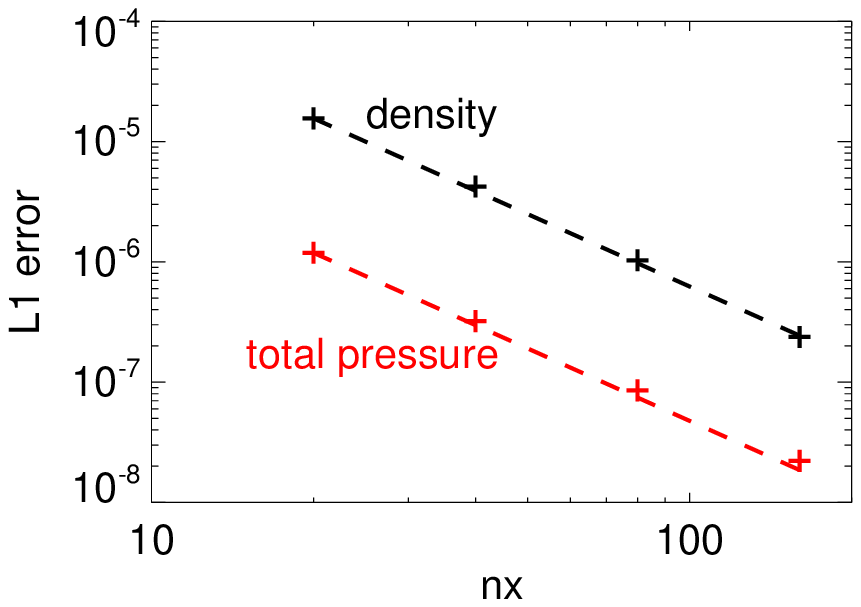}}
 \caption{Error on the density (black) and total pressure (red) after one orbit of evolution for the magnetic shearing wave. }
             \label{fig:test_shear_magnetic_error}%
\end{figure}

We have performed two-dimensional simulations (in the $x$-$y$ plane)
of both types of waves, with fiducial parameters $L_x = 1$, $L_y=4$,
$A=10^{-3}$, $n_y = 1$, $\gamma=1.4$, $\rho_0=1$, $c_{s0}=1$,
$\Omega=1$.
 We used various radial resolutions, $\Delta=
 1/20,\,1/40,\,1/80,\,1/160$ with the azimuthal grid size twice the
 radial in each case. Fig.~\ref{fig:test_shear_image} shows the distribution of sound speed one orbit after the beginning of the simulation of the magnetic shearing wave with $n=2$. Fig.~\ref{fig:test_shear_magnetic_error} shows the L1 norm of the deviation from the analytical solution for the density (black) and the total pressure (red) as a function of resolution. The convergence is quadratic as expected for a second order code. In the case of the entropy wave, we obtain a similar convergence of the error on the density distribution. Pressure and velocity perturbations are induced only at truncation error, likely because the pressure, and therefore energy $E^\prime$, is uniform in the box.

\subsection{Epicyclic oscillations}
A basic test of the implementation of the shearing box source terms is
epicyclic oscillations. We initiate epicyclic oscillations of a
shearing box with uniform density and pressure by setting the radial
velocity to a uniform and constant value $v_0=c_s$. For the momentum evolution, we obtain very similar results to Stone \& Gardiner (2010): the energy of the epicyclic oscillations $E_{\rm epi}=0.5\rho\left(v_x^2 + 4v_y^{\prime2} \right)$ is conserved to truncation error, as expected since we use the same Crank-Nicholson algorithm, while a small dispersion error is observed at low resolution. The pressure remains at its initial value, up to truncation error, which shows that the source term in the energy equation involving the Reynolds stress is accurately computed. 

We also considered epicyclic oscillations in a radially non-uniform box, containing a radially varying entropy wave with uniform pressure as
\begin{eqnarray}
\rho &=& \rho_0\left\lbrack 1 + A\cos \left(\frac{2\pi n_x x}{L_x}\right) \right\rbrack, \\
P &=& P_0, 
\end{eqnarray}
where $n_x$ is the number of wavelengths in the radial extent of the
box. We performed one dimensional simulations with the following
parameters : $L_x=1$, $v_x=c_{s0}$, $c_{s0}=1$, $A=10^{-3}$ and a number of
grid points varying between 20 and 160. In this case, numerical errors
induce pressure perturbations, which converge quadratically.

\section{Robustness  of results on numerical parameters and set up}
\label{sec::ic}

In this appendix we present some ancillary simulations to Sections 4
and 5
showing the robustness of the main results. In the main body of the
text we examine the effect of resolution and explicit diffusion. Here
we test initial conditions and confirm that our choice of resolution
is adequate when using explicit diffusion coefficients. 

\subsection{Initial conditions}

\begin{figure}
\centering
\vspace{-1mm}
\scalebox{0.5}{\includegraphics{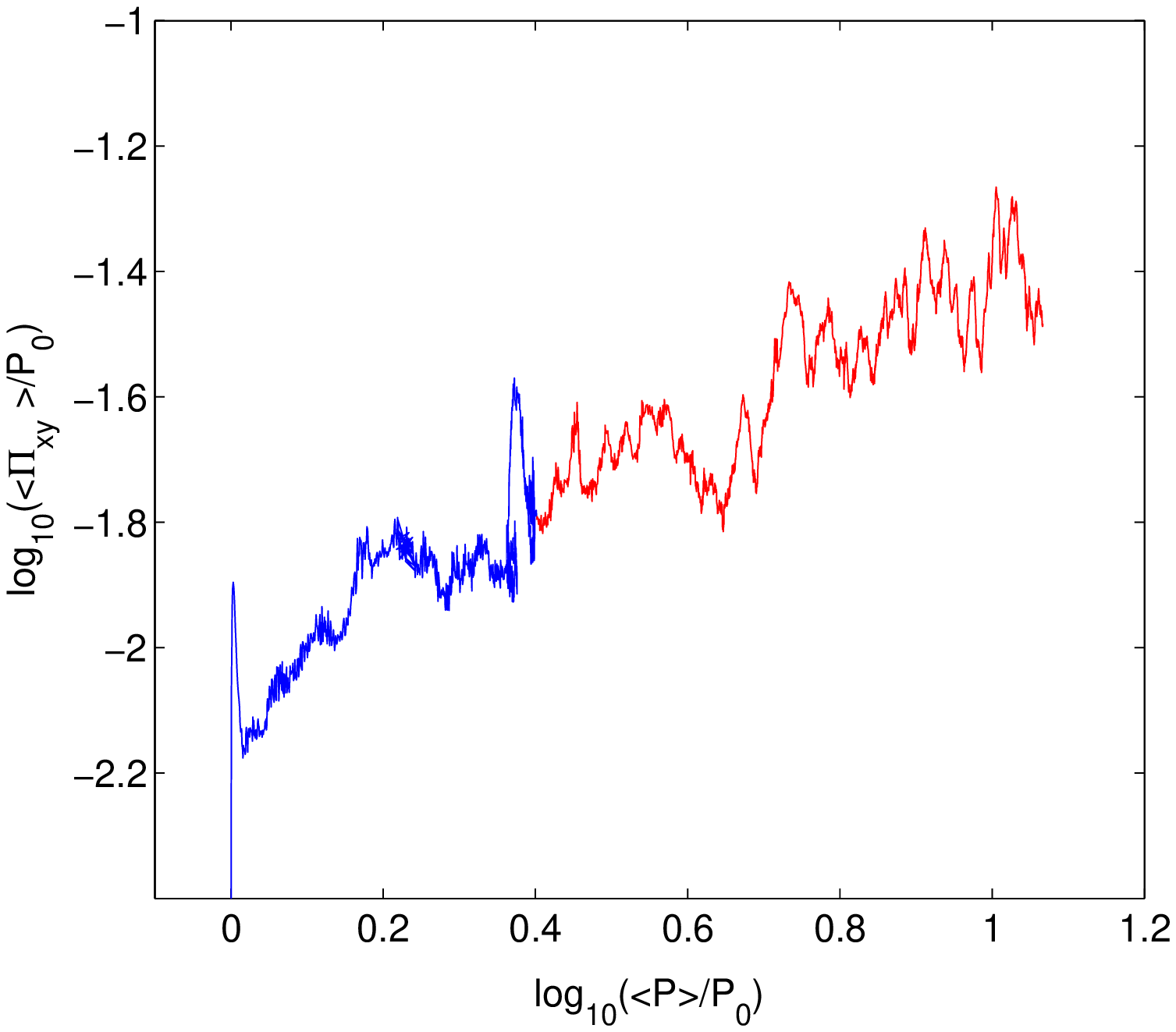}}
 \caption{The time evolution of the Maxwell stress normalized by the
  initial gas pressure, as a function of the volume averaged pressure
  for $L=2H_{0}$ simulations. The blue curve is from the simulation bringing the system to a thermal equilibrium and the red curve is when the cooling is removed.}
             \label{fig:SvPeq}%
\end{figure}

In Section 4 we consider zero-net-flux systems that are out of
equilibrium and evolving with time. The phenomena that we are
primarily interested in occur relatively early in the simulations and
so the initial conditions may not have been completely forgotten and
may be influencing our results. We check how robust they are to the
choice of initial condition. 
Instead of small amplitude noise we use as an initial condition a
turbulent quasi-equilibrium. To generate this quasi-equilibrium we used the same computational set-up as the fiducial $L=2H_{0}$ zero-net-flux simulation but we introduce a cooling law $\Lambda=\theta P^{2}$ into the energy equation. This prescription ensured a stable non-zero thermal equilibrium point from which we could restart the simulation with out any cooling. The results are shown in Fig.~\ref{fig:SvPeq} and show the increase in stress with pressure with a comparable $q$ to our fiducial simulation, $q=0.35 - 0.65$. 

\subsection{Dissipation}

In Section 4.1.2 we perform simulations with explicit diffusion and
resistivity 
accounted for with $Re=1250$ and $R_m=5000$ and with a grid of
$\Delta=1/64$. 
It is necessary to check
that 
the grid is sufficient small to ensure negligible numerical diffusion
and dissipation. Our approach was to consider the contributions to the
volume 
averaged internal energy $\langle\epsilon\rangle$. 
The rate of change in $\langle\epsilon\rangle$ can be written as
\begin{align}
\langle\dot{\epsilon}\rangle
&=-\langle P \nabla \cdot \bold{v}\rangle + \langle D_\text{phy}\rangle+\langle D_\text{num} \rangle
\end{align}  
where $\langle D_\text{phy}\rangle$ and $\langle D_\text{num}\rangle$ are the volume averaged contributions from physical and numerical dissipations respectively.

\begin{figure}
\centering
\scalebox{0.5}{ \includegraphics{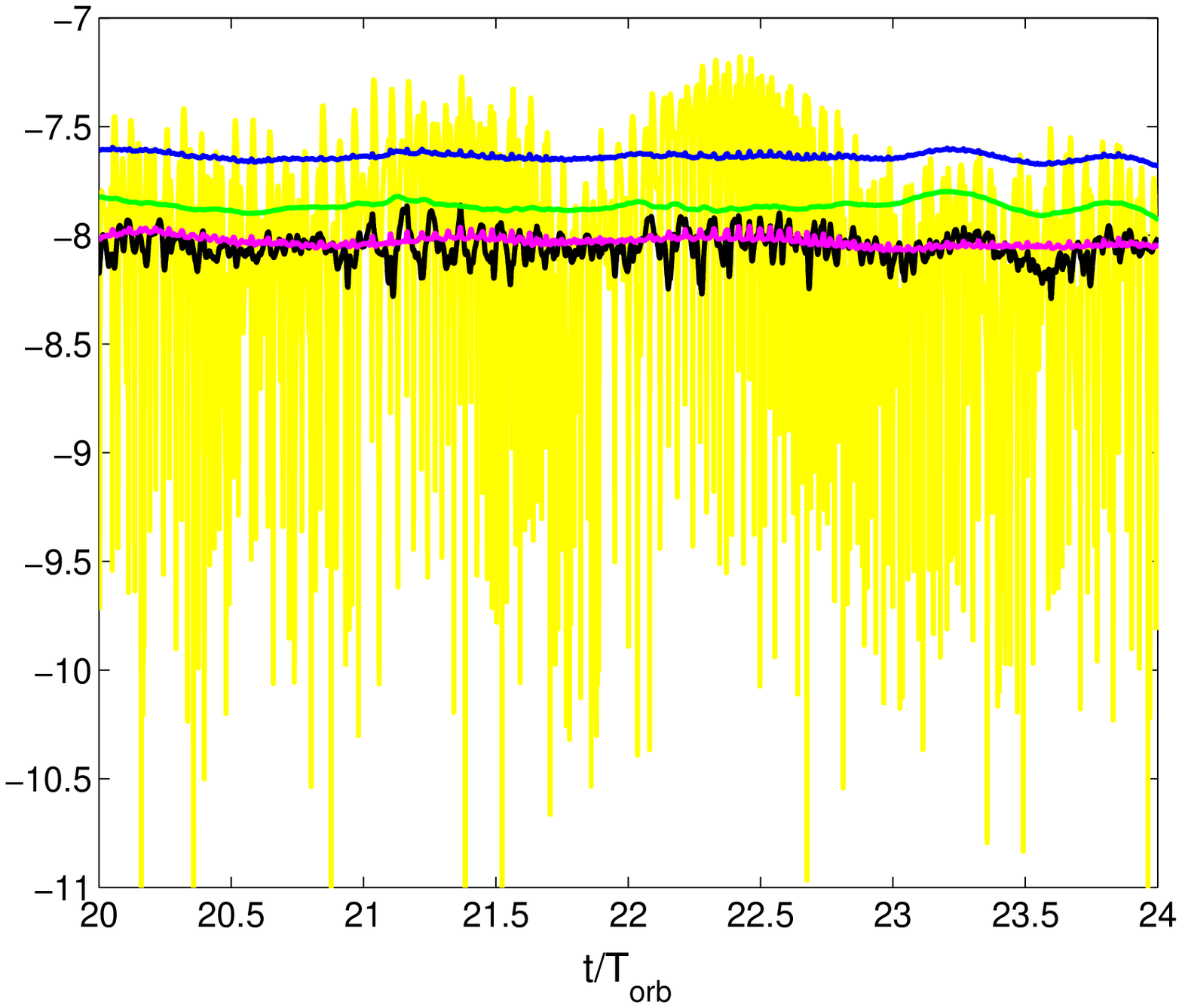}}
\caption{Various box-averaged thermal energy input terms plotted on a $log_{10}$ scale over 4 orbits. The green curve is the Ohmic heating, pink is the viscous heating, blue is the sum of the Ohmic and the viscous heating, yellow is the pressure heating term and the black curve is the numerical heating.}
\label{Fig::energyinput}%
\end{figure}

For our choice of parameters to be appropriate we must have $\langle
D_\text{num}\rangle < \langle D_\text{phy} \rangle$. We compute these
quantities for a $L=H_{0}$ zero net-flux box with the above choice of
parameters. These box averaged quantities are plotted in
Fig.~\ref{Fig::energyinput}. This shows that the total time averaged numerical
diffusion is significantly less than the total physical dissipation
but roughly equal to the (subdominant) viscous dissipation.

To further check that a resolution of $\Delta=1/64$ gives appropriate
results for our chosen $Re$ and $R_m$, we
undertook $L=2H_0$ heating simulations with $\Delta=1/64$ and
$1/128$. Both yielded $q\approx 0.5$ and a maximum stress that was
roughly consistent. 

\end{appendix}

\end{document}